\documentclass[12pt, a4paper]{article}
\usepackage{graphicx}

\usepackage{amsmath,amsfonts,latexsym,mathrsfs}
\usepackage{natbib}
\usepackage{color}
\usepackage[usenames,dvipsnames]{xcolor}
\usepackage{graphicx}

\usepackage{accents}
\newlength{\dhatheight}

   \makeatletter
\def\@tempa#1{\@xp\@tempb\meaning#1\@nil#1}
\def\@tempb#1>#2#3 #4\@nil#5{%
  \@xp\ifx\csname#3\endcsname\mathaccent
    \@tempc#4?"7777\@nil#5%
  \else
    \PackageWarningNoLine{amsmath}{%
      Unable to redefine math accent \string#5}%
  \fi
}
\def\@tempc#1"#2#3#4#5#6\@nil#7{%
  \chardef\@tempd="#3\relax\set@mathaccent\@tempd{#7}{#2}{#4#5}}

\@tempa\widehat
\makeatother

\definecolor{bubblegum}{rgb}{0.99, 0.76, 0.8}
\definecolor{carnationpink}{rgb}{1.0, 0.65, 0.79}
\definecolor{cherryblossompink}{rgb}{1.0, 0.72, 0.77}
\definecolor{classicrose}{rgb}{0.98, 0.8, 0.91}
\definecolor{cottoncandy}{rgb}{1.0, 0.74, 0.85}
\definecolor{electriclavender}{rgb}{0.96, 0.73, 1.0}
\definecolor{flamingopink}{rgb}{0.99, 0.56, 0.67}
\definecolor{hotpink}{rgb}{1.0, 0.41, 0.71}
\definecolor{ticklemepink}{rgb}{0.99, 0.54, 0.67}

\definecolor{KMpink}{RGB}{255, 153, 255}
\definecolor{KMgreen}{RGB}{0, 127, 0}

\begin{document}

\baselineskip=15pt
\vspace{-2.5cm}
\title {Radiating solitary waves in  coupled Boussinesq equations}


\date{}
\maketitle
\vspace{-22mm}
\begin{center}
{\bf R.H.J. Grimshaw, K.R. Khusnutdinova and
K.R. Moore} \\[2ex]
Department of Mathematical Sciences, Loughborough University, \\
Loughborough LE11 3TU, UK\\ 
\vspace{4mm}

\end{center}

\abstract{
In this paper we are consider
radiating solitary wave solutions of  coupled regularised Boussinesq equations. 
This type of solution consists of a  leading solitary wave with a small-amplitude co-propagating oscillatory tail, 
and emerges from a pure solitary wave solution of a symmetric reduction of the full system. 
We construct an asymptotic solution, where the leading order approximation in both components is 
obtained as a particular solution of the regularised Boussinesq equations in the symmetric case. 
At the next order, the system uncouples into two linear non-homogeneous 
ordinary differential equations with variable coefficients, 
one correcting the localised part of the solution,  which we find analytically, and the other describing 
the co-propagating oscillatory tail. This  latter equation is a fourth-order ordinary differential equation 
and is solved approximately by two different methods, each exploiting the assumption that the 
leading solitary wave has a small amplitude, and thus enabling an explicit estimate for the 
amplitude of the oscillating tail.
These  estimates
are compared with 
corresponding numerical simulations.
\bigskip
\bigskip
\bigskip

\bigskip

{\bf Corresponding author:} K.Khusnutdinova@lboro.ac.uk

\newpage

\section{Introduction}  \label{sec:sol_waves}

In the last   couple of decades it has been shown that certain Boussinesq-type equations can be used to model the propagation of
 long nonlinear longitudinal bulk strain waves in  many elastic waveguides, including for example rods and plates, see   \citet{Samsonov01, Porubov03}.
 The predicted longitudinal bulk strain solitary waves were observed in experiments by \citet{Dreiden88, Samsonov98, Semenova05}.  Recently, these theoretical and experimental studies were extended to some types of adhesively bonded layered bars by \citet{KS,KSZ,DSSK,Dreiden12}. In particular, it was shown that nonlinear longitudinal bulk strain waves in a bi-layer with a sufficiently soft adhesive bonding can be modelled by a system of coupled regularised Boussinesq (cRB) equations:
\begin{eqnarray}
f_{tt} - f_{xx} &=& \frac 12 (f^2)_{xx} + f_{ttxx} - \delta (f-g)\,, \nonumber \\
g_{tt} - c^2 g_{xx} &=& \frac 12  \alpha  (g^2)_{xx} + \beta  g_{ttxx} + \gamma  (f-g)\,. \qquad
\label{fg}
\end{eqnarray}
The equations are written in non-dimensional form, $f$ and $g$ describe the  longitudinal strains in the layers, while the coefficients are defined by the physical and geometrical parameters of the problem (see \citet{KSZ} for details). 
%

In the symmetric case, when $c=\alpha=\beta=1$, system (\ref{fg}) admits a reduction $g=f$, where $f$ satisfies the single   regularised Boussinesq equation
\begin{equation}
f_{tt} - f_{xx} = \frac 12 (f^2)_{xx} + f_{ttxx}\,.
\label{f_intro}
\end{equation}
The Boussinesq equation (\ref{f_intro}) has particular solutions in the form of pure solitary waves:
\begin{equation}
f = a\  {\rm sech}^2 \frac{x - v t}{\Lambda}\,, \quad  a= 3 (v^2 - 1)\,, 
\quad  \Lambda = \frac{2 v}{\sqrt{v^2 - 1}} \,.
\label{soliton_intro}
\end{equation} 
Figure \ref{pure_solitary_wave} illustrates the pure solitary wave solution (\ref{soliton_intro}) for two 
values of the parameter $v$.

\begin{figure}[h]
\begin{center}
\includegraphics[width=6cm]{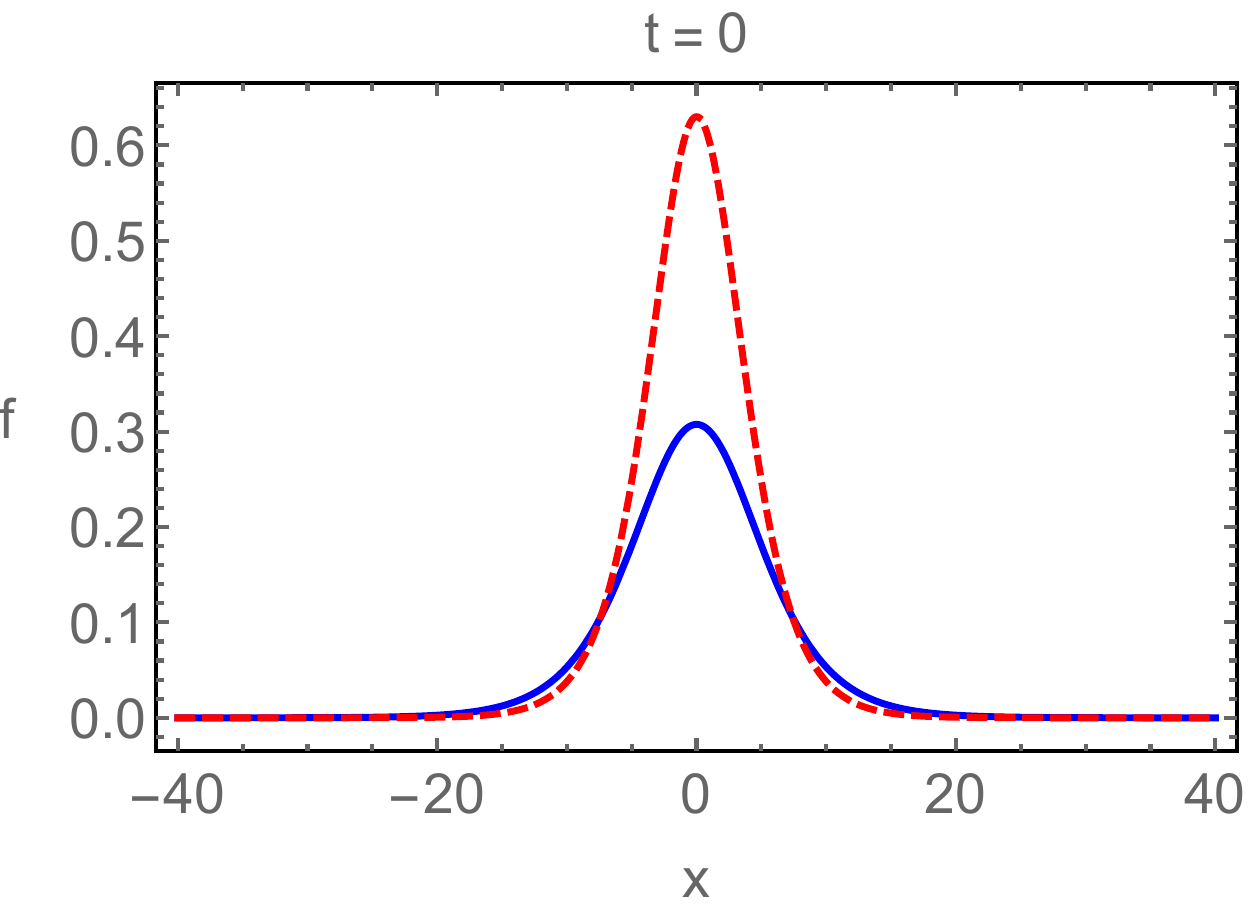} 
\caption{\small Pure solitary wave solution (\ref{soliton_intro}) for $v=1.05$ (solid line) and $v = 1.1$ (dashed line).}
\label{pure_solitary_wave}
\end{center}
\end{figure}

However, in the cRB system of equations (\ref{fg}), when the value of $c$ is close to 1,
 that is,  the characteristic speeds of the linear waves in the layers are close,  these 
 ``pure'' or ``classical''  solitary wave solutions, rapidly decaying to zero in their tail regions, 
 are structurally unstable and are replaced with ``radiating'' solitary waves (see \citet{KSZ,KM}), 
 that is a solitary wave radiating a co-propagating one-sided oscillatory tail, using  the terminology in \citet{BGK, Shrira, Bona}. A typical illustration of radiating solitary wave solutions in each component of the cRB equations (\ref{fg}) is depicted in Figure \ref{RSW_ch2}.

\begin{figure}[h]
\begin{center}
\includegraphics[width=4.8cm]{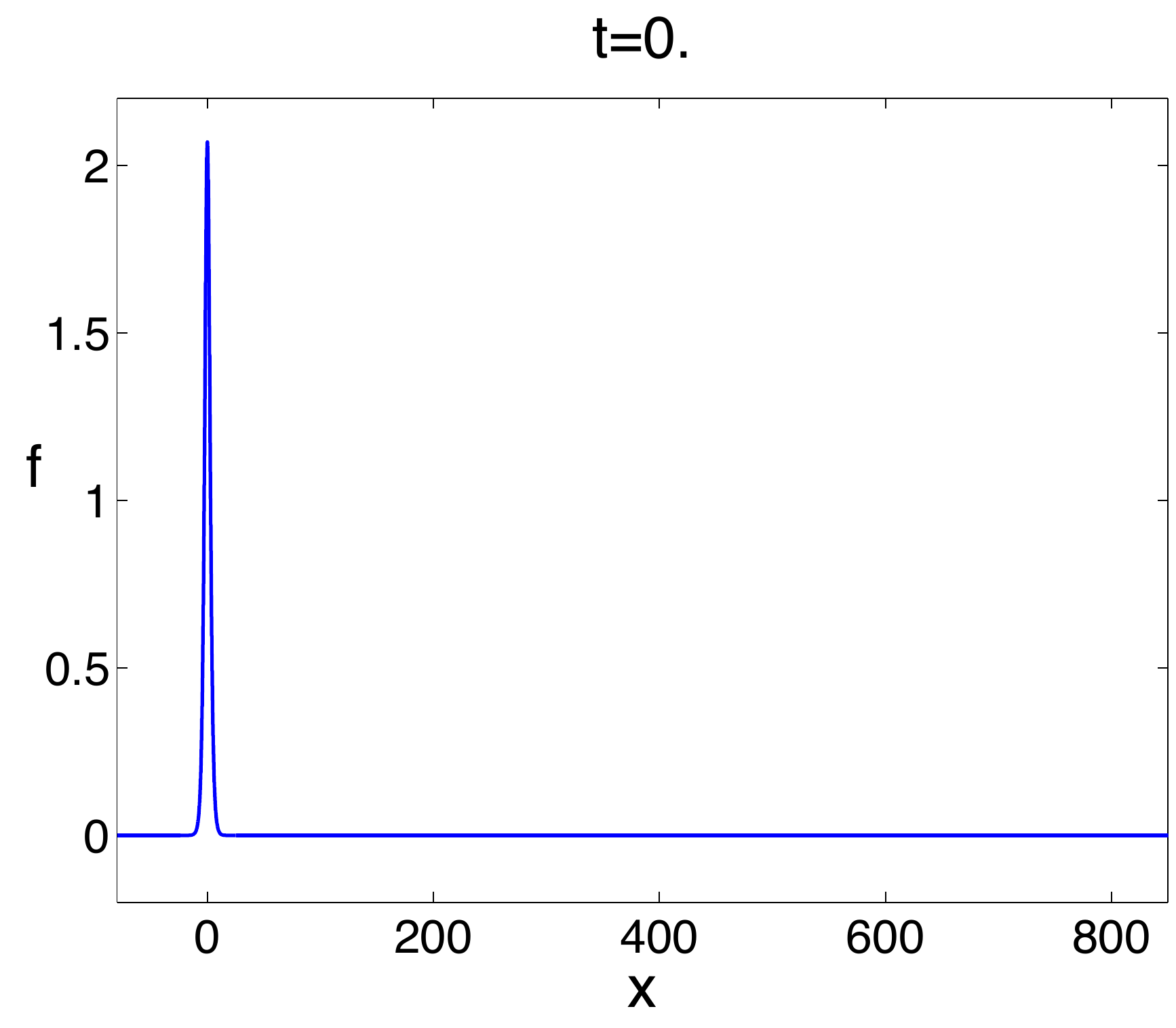} \quad 
\includegraphics[width=4.8cm]{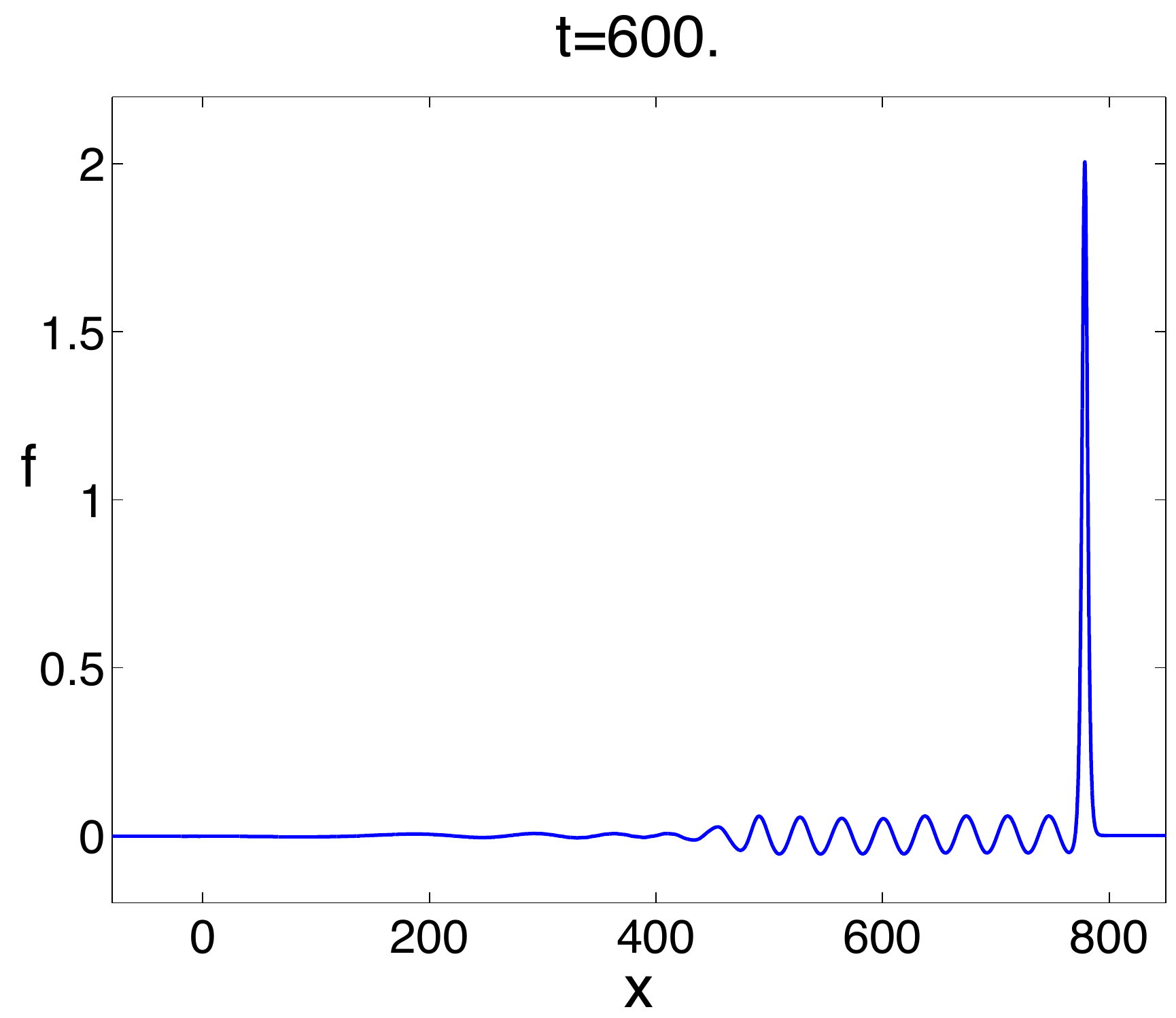} \quad \\
\includegraphics[width=4.8cm]{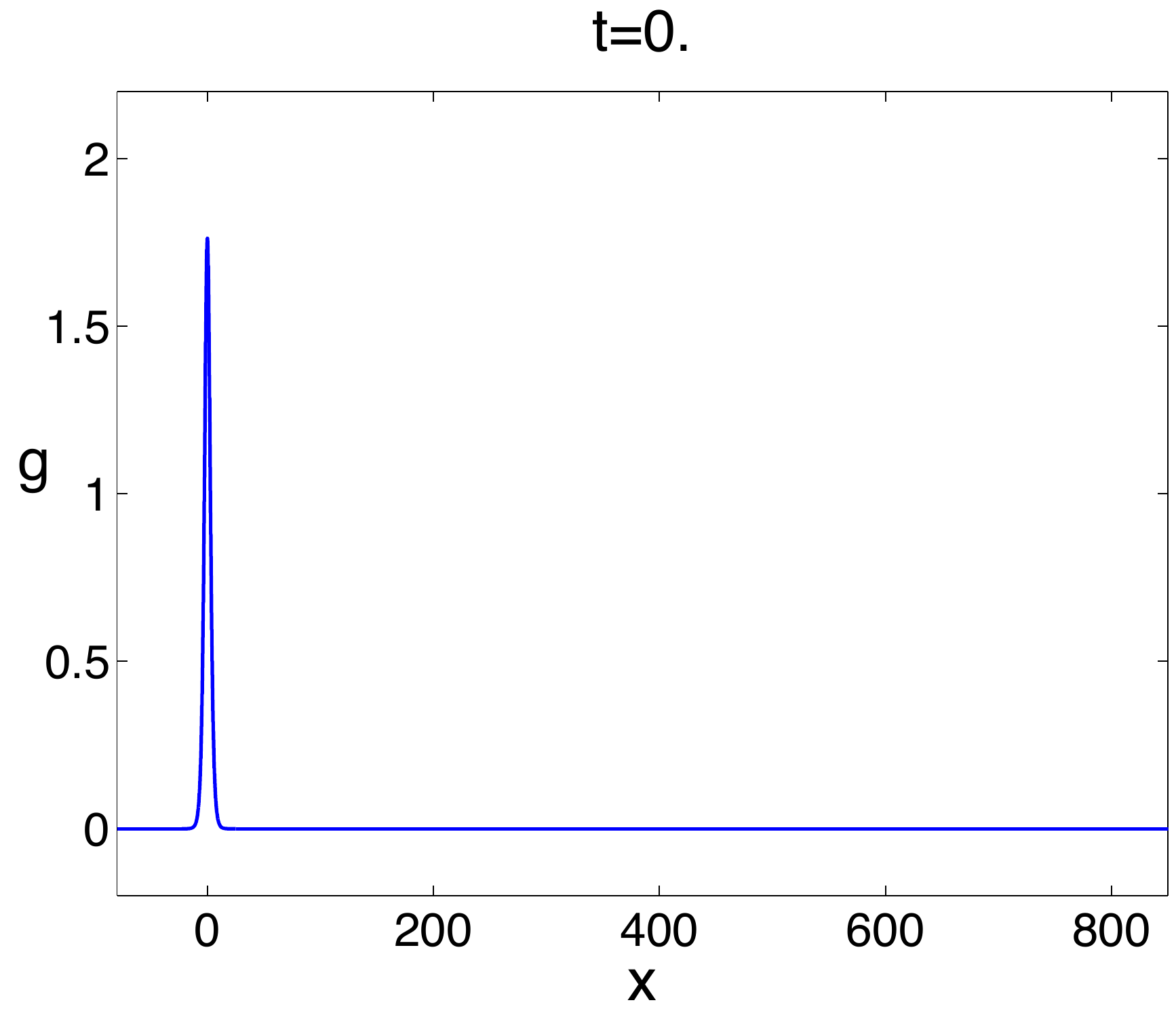} \quad 
\includegraphics[width=4.8cm]{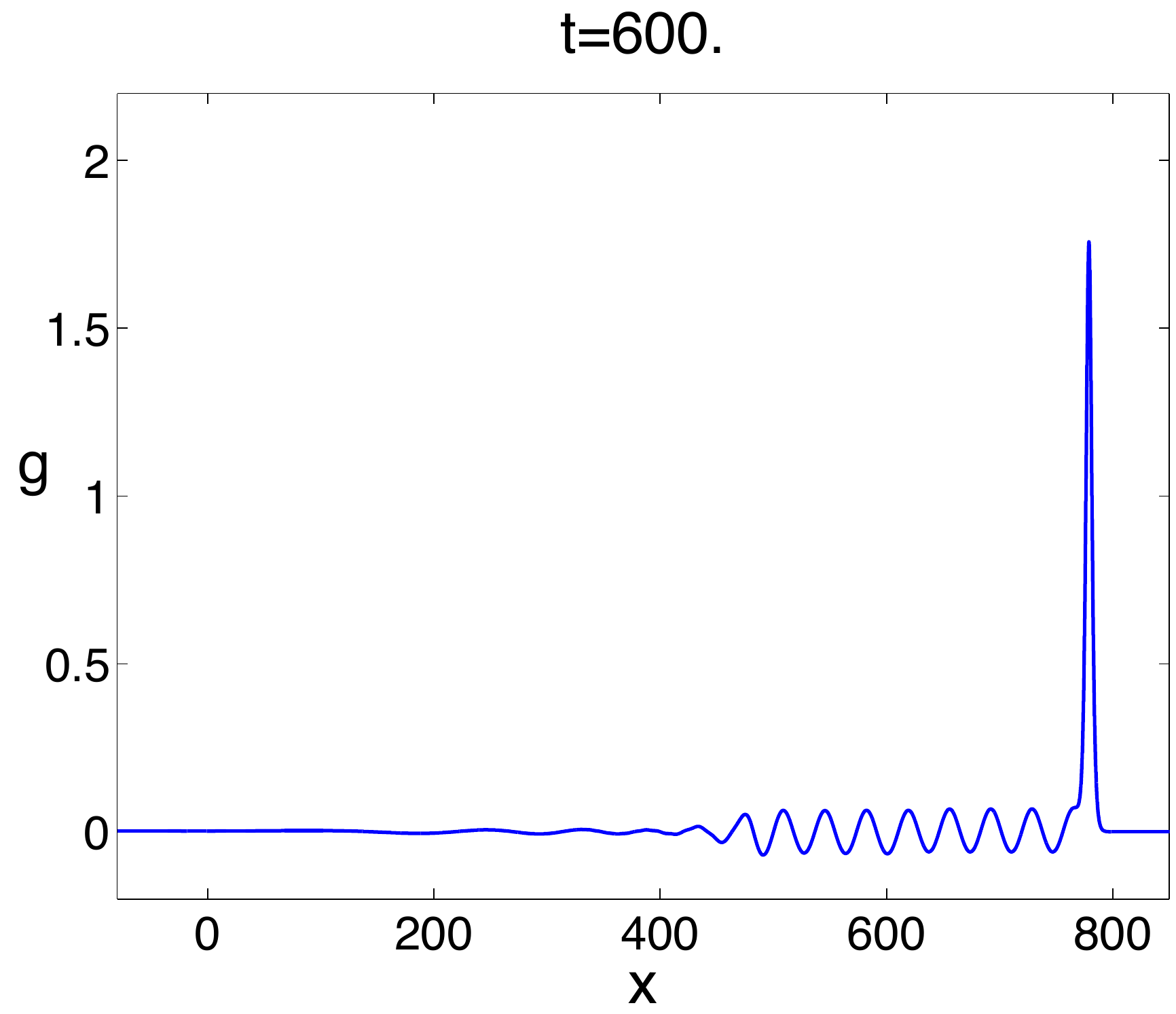} 
\caption{\small Typical generation of radiating solitary waves in the system of cRB equations (\ref{fg}), from pure solitary wave initial conditions. Here $c=1.05$, $\alpha=\beta=1$, $\gamma=\delta=0.01$; $v=1.3.$}
\label{RSW_ch2}
\end{center}
\end{figure}

There have been extensive studies of generalised and radiating solitary waves, especially in the context of fluid mechanics,
see for instance,  \citet{Boyd,Grimshaw,GI,GJ,Karpman,Lombardi,VB}). 
The most commonly studied systems supporting these non-local nonlinear long waves include perturbed KdV equations; coupled KdV systems; perturbed NLS equations and coupled NLS systems. 
The underlying reason for the occurrence of generalised and radiating solitary waves is due to a resonance between a long wave (with wavenumber $k\approx 0$) and a short wave (with some finite wavenumber). Steady generalised solitary waves are necessarily symmetric, and they support oscillating tails on both sides of the localised central core.
However this means they usually cannot be realised physically since the group velocity of the oscillating tails is the same at both ends. In practice, one instead finds that radiating solitary waves are generated, that is asymmetric non-local solitary waves with the oscillating tail appearing on only that side of the central core determined by the outgoing group velocity.
Strictly, such radiating solitary waves must be unsteady, and decay due to the emitted radiation. But if the amplitude of this radiating tail is sufficiently small, then this rate of decay is sufficiently small that a quasi-steady assumption can be made.

In the system  (\ref{fg}) considered here, long-wave oscillations  are radiated 
from the central localised core, due to the nature of the coupling terms in the equations and the resulting structure of the linear dispersion relation.
The linear dispersion relation for (\ref{fg}) was analysed in \citet{KSZ} by assuming that the 
coefficients in (\ref{fg}) are perturbed compared to the symmetric case, but remain positive. 
The dispersion relation has the form
\begin{equation}
[k^2 (1-p^2) - k^4 p^2 + \delta] [k^2 (c^2-p^2) - \beta k^4 p^2 + \gamma] = \gamma \delta,
\label{disp}
\end{equation}
where $k$ is the wavenumber and $p$ is the phase speed. A typical plot of (\ref{disp})
is  shown in Figure \ref{disp_curves_intro}(a). A significant difference with the linear dispersion curve of the reduction (\ref{f_intro})  consists in the appearance of the second (upper) branch, going to infinity as $k \to 0$, and approaching zero, remaining above the lower branch, as $k \to \infty$.

\begin{figure}[h]
\begin{center}$
\begin{array}{cc}
\includegraphics[width=5.3cm]{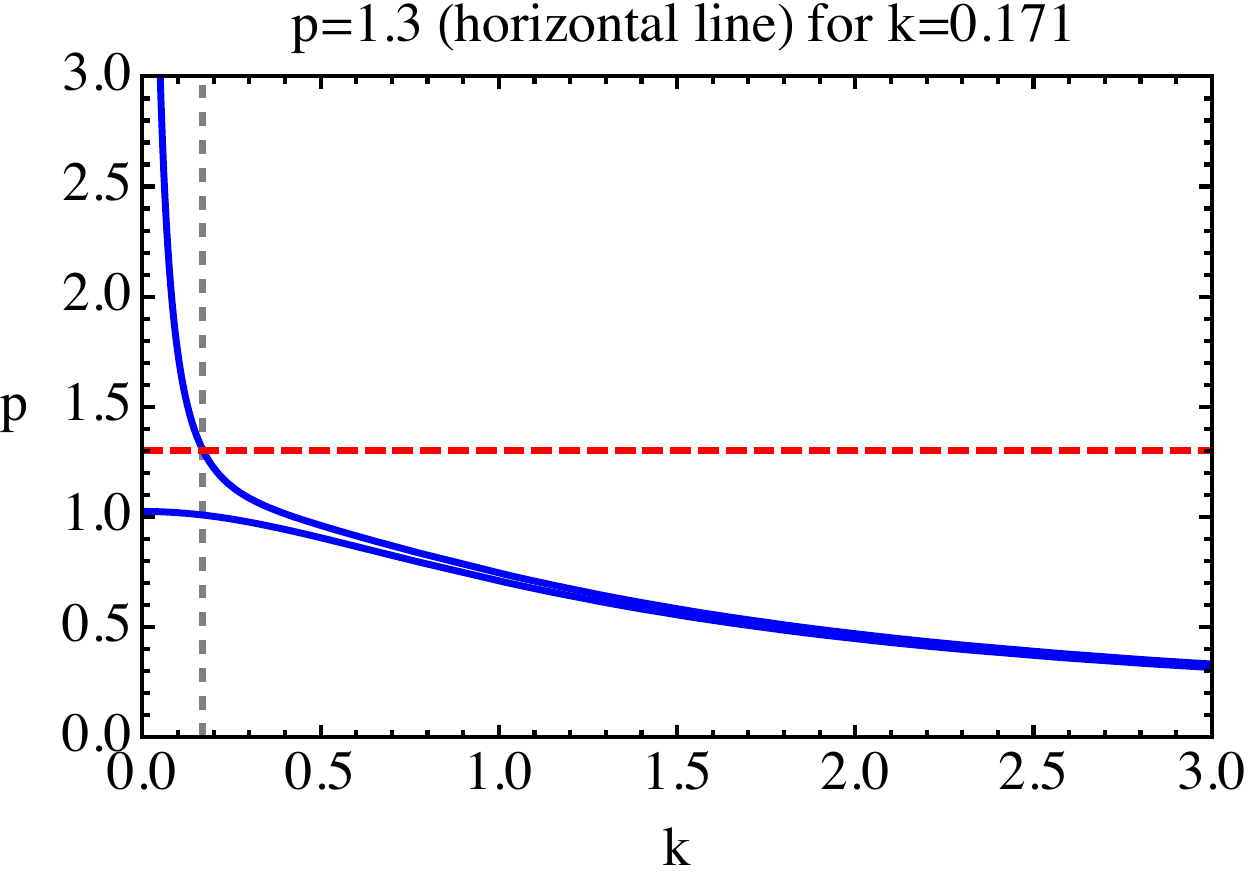} \ \ \ \ \ &
\includegraphics[width=5.3cm]{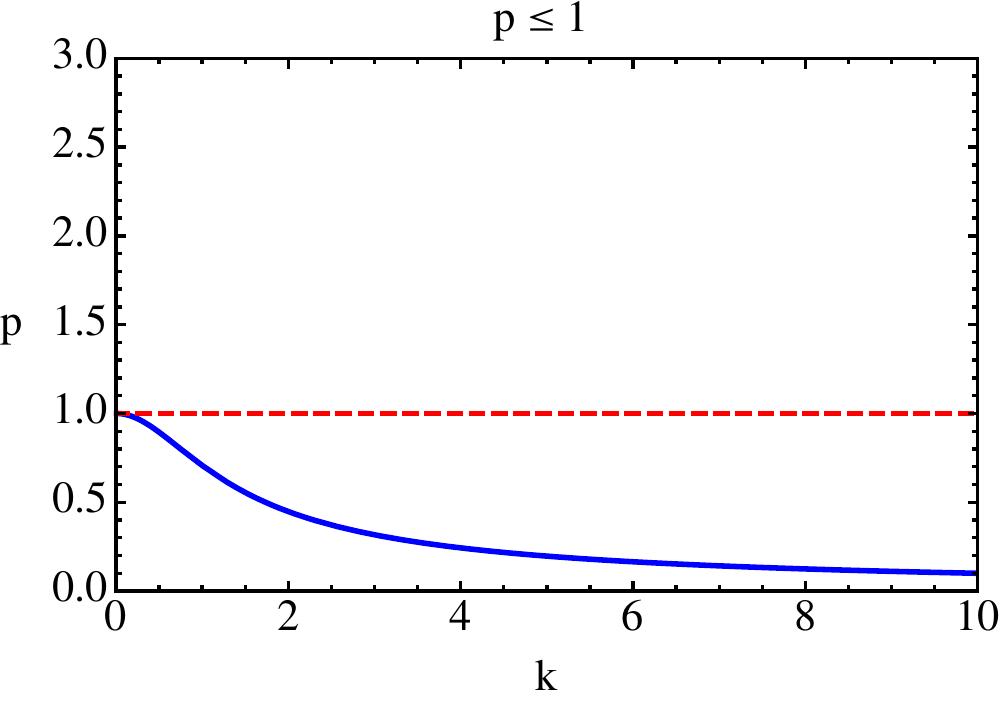}  \\
 {\rm (a)} \ \ \ \ \  & {\rm (b)}  
\end{array}$
\caption{\small (a) Two branches of the linear dispersion curve of the cRB equations (\ref{fg}) for $c=1.05, \beta = 1,  \delta = \gamma =0.01$ and illustration of a possible resonance for $p= 1.3$ (horizontal line), and (b) Linear dispersion curve of the reduction (\ref{f_intro})  in the symmetric case $c=1, \beta = 1,  \delta = \gamma =0.01$.}
\label{disp_curves_intro}
\end{center}
\end{figure}

The classical or pure solitary waves of the single Boussinesq equation (\ref{f_intro})  arise as a bifurcation from wavenumber $k = 0$ of the linear wave spectrum, shown in Figure \ref{disp_curves_intro}(b), when there is no possible resonance with any linear wave for any value of $k$. The solitary wave speed $v$  is greater than  the linear long wave speed, that is,  $v > 1$, while the speed of a linear wave of any wavenumber is smaller, that is $p \le 1$.  This becomes impossible when the symmetry is broken. Instead, radiating solitary waves  arise  for the case when there is a possible resonance with the upper branch for some finite non-zero value of $k$. For example, a possible resonance is shown in Figure \ref{disp_curves_intro}(a) for $v=p=1.3$.  
Recently, radiating solitary waves have been experimentally observed in two- and three-layered elastic waveguides with soft bonding layers by \citet{DSSK}.
The purpose of this paper is to find explicit expressions for the radiating waves in the case when the symmetry 
is only weakly broken. Our strategy is to expand the coefficients in (\ref{fg}) about the symmetric case
$c=\alpha=\beta=1$ using a perturbation parameter $\mu \ll 1$. This yields a set of linear equations at 
$O(\mu )$ described in section 2. These equations are in effect a form of linearisation about the solitary wave
(\ref{soliton_intro}). They can be decoupled into two equations, one describing the localised perturbation of the solitary wave, 
developed in section 2.1 and capable of an explicit solution, and the other describing the 
radiating solitary waves, developed in section 2.2. This latter system is fourth order, and here we exploit the 
solitary wave amplitude parameter in (\ref{soliton_intro}) to obtain approximate solutions in the limit when
this amplitude is small. These radiating solitary wave solutions  of the cRB equations are then compared with   some 
relevant  numerical simulations in section 3. We conclude in section 4.

\section{Asymptotic perturbation from the symmetric case}
It is now convenient to put the  cRB equations (\ref{fg}) in the form,%
\begin{eqnarray}
f_{tt} - f_{xx} &=& \frac 12 (f^2)_{xx} + f_{ttxx} - \delta (f-g), \nonumber \\
g_{tt} - g_{xx} &=& \frac 12   (g^2)_{xx} +  g_{ttxx} + \gamma  (f-g)  + \
\mu \left[ Ag + \frac 12 B g^2 + Cg_{tt}    \right]_{xx},
\label{eqn:perturb_bous_sys}
\end{eqnarray}
where  $(c^2-1)= A\mu$, $(\alpha-1)= B\mu $, $(\beta - 1)= C\mu$, 
where $A, B, C=O(1)$ and  $|\mu| \ll 1$ is a small parameter.
Seeking travelling wave solutions of the form $f =f(\xi )$, $g=g(\xi )$ for $\xi = x -vt$, we have from (\ref{eqn:perturb_bous_sys})
\begin{eqnarray}
(v^2 - 1)f_{\xi \xi } &=& \frac{1}{2} (f^2)_{\xi \xi } + v^2 f_{\xi \xi \xi \xi } - \delta (f-g), \nonumber \\
(v^2 - 1)g_{\xi \xi } &=& \frac{1}{2}(g^2)_{\xi \xi } +  v^2 g_{\xi \xi \xi \xi } + \gamma  (f-g)  
+ \mu \left[ Ag + \frac 12 B g^2 + v^2 C g_{\xi \xi }    \right]_{\xi \xi }.
\label{eqn:perturb_bous_sys2}
\end{eqnarray}
We look for a solution of (\ref{eqn:perturb_bous_sys2}) in the form of the following asymptotic expansions:
\begin{eqnarray}
f = f_0+\mu f_1 + O(\mu^2) \,, \quad 
g = g_0 + \mu g_1+ O(\mu^2) ,  \quad v^2 =v^{2}_0 (1 + \mu v_1 + O(\mu^2) ) \,. 
\label{eqn:perturb_exp}
\end{eqnarray} 
Substituting (\ref{eqn:perturb_exp}) into (\ref{eqn:perturb_bous_sys2}), 
at the leading order we recover the particular solution 
(\ref{soliton_intro}) of the single regularised Boussinesq equation, as discussed above in section \ref{sec:sol_waves}, 
\begin{eqnarray}
f_0 = g_0 = a_0\  {\rm sech}^2 \left(\varepsilon \xi /2 \right), \quad \mbox{where} \quad 
\varepsilon = \frac{\sqrt{v_{0}^2 - 1}}{v_0} \,, \quad v_{0}^{2} = \frac{1}{1- \varepsilon^2 } \,, \quad 
a_0 = 3 (v_{0}^2 - 1) = \frac{3 \varepsilon^2 }{1 - \varepsilon^2 } \,.
\label{eqn:soliton_later}
\end{eqnarray}
At the next order  $O(\mu)$ we find that, 
\begin{eqnarray}
(v_{0}^2 - 1)f_{1\xi \xi } &=&  (f_0 f_1 )_{\xi \xi } + v_{0}^2 f_{1\xi \xi \xi \xi } - \delta (f_1-g_1 )
 + v_0^2v_1( f_{0\xi \xi }-f_0 )_{\xi \xi}, \nonumber \\
(v_{0}^2 - 1)g_{1\xi \xi } &=& (f_0 g_1)_{\xi \xi } +  v_{0}^2 g_{1\xi \xi \xi \xi } + \gamma  (f_1-g_1 )  
+ v_0^2v_1( f_{0\xi \xi }-f_0 )_{\xi \xi} + \nonumber  \\
& &  \left[ Af_0 +  \frac 12 B f_{0}^2 + v_{0}^2 C f_{0\xi \xi }    \right]_{\xi \xi } \,. 
\label{eqn:perturb_bous_sys_Omu}
\end{eqnarray}
Introducing the variables $\phi=f_1-g_1$ and $\psi=f_1+(\delta/\gamma)g_1$, the system 
(\ref{eqn:perturb_bous_sys_Omu}) becomes 
\begin{eqnarray}
(v^{2}_0 -1 ) \phi_{\xi \xi }  &=&  (f_0\phi)_{\xi \xi } + v^{2}_0 \phi_{\xi \xi \xi \xi } - (\delta+\gamma)\phi - P(f_0)_{\xi \xi }, \label{eqn:perturb_sys_phi_xi} \\
(v^{2}_0 - 1) \psi_{\xi \xi } &=&  (f_0\psi)_{\xi \xi } + v^{2}_0 \psi_{\xi \xi \xi \xi } + 
\frac{\delta}{\gamma} P(f_0)_{\xi \xi  } + v_0^2v_1 (1 +\frac{\delta }{\gamma})( f_{0\xi \xi }-f_0 )_{\xi \xi}, 
\label{eqn:perturb_sys_psi_xi}  \\
\hbox{where} \quad P(f_0) & = &  Af_0 +  \frac 12 B f_{0}^2 + v_{0}^2 C f_{0\xi \xi }  
=  [A + (v_{0}^2 -1)C]f_{0} + \frac 12 (B-C) f_{0}^2 \,. \label{P}
\end{eqnarray}
The system is uncoupled and so equations  (\ref{eqn:perturb_sys_phi_xi}), (\ref{eqn:perturb_sys_psi_xi}) 
can respectively each be solved for $\phi , \psi$ alone.
\subsection{The localised part of the solution}
 \label{sec:RSW_non_osc_phi}

First we consider the equation (\ref{eqn:perturb_sys_psi_xi}) for $\psi $.
Introducing a change of variable  and integrating twice yields
\begin{eqnarray}
\psi_{\chi\chi}  +  4(3{\rm sech}^2\chi -1) \psi = F(\chi ) + C_1\chi + C_2 \,, \quad
\chi = \frac{\varepsilon}{2 }\xi  
\label{eqn:perturb_sys_psi_chi} \,,
\end{eqnarray}
where 
$C_{1,2}$ are integration constants, and 
\begin{eqnarray*} 
F=  -\frac{4\delta}{\gamma(v_{0}^2-1)} P(f_0) + \frac{2v_1(1 + \frac{\delta }{\gamma })}{v_0^2-1}[2f_0 + f_0^2].
\end{eqnarray*}
We set the constants of integration $C_1, C_2=0$ since we seek solutions for  $\psi$ which are localised as 
$\chi \rightarrow \pm \infty$.
This second order inhomogeneous ordinary differential equation can be solved using the method
of variation of parameters. The homogeneous part of (\ref{eqn:perturb_sys_psi_chi}) is
\begin{eqnarray}
\psi_{\chi\chi}  +  4(3{\rm sech}^2\chi -1) \psi = 0,
\label{eqn:perturb_sys_psi_chi_HE}
\end{eqnarray}
and has a localised  solution of the form
\begin{eqnarray}
\psi_1(\chi) =  {\rm sech}^2(\chi)\, {\rm tanh}(\chi)  \,,
\label{eqn:psi_1}
\end{eqnarray}
which is proportional to $f_{0\chi }$, since (\ref{eqn:perturb_sys_psi_chi_HE}) is the 
linearisation of the solitary wave equation for $f_{0}$. 
A second linearly independent solution $\psi_2$ of (\ref{eqn:perturb_sys_psi_chi_HE})
 can be found using  the Wronskian 
\begin{eqnarray}
\psi_1\psi_{2\chi } - \psi_2 \psi_{1\chi } = W,  
\label{eqn:abels}
\end{eqnarray}
where $W$ is a constant. Solving (\ref{eqn:abels}) for $\psi_2 $ yields
\begin{eqnarray}
\psi_2(\chi) = \frac{W}{32}\Big[ 60\chi - 32\ {\rm coth}(\chi) +  16  \ {\rm sinh}(2\chi)  +  {\rm sinh}(4\chi)\Big]  
\,{\rm sech}^2(\chi) \,{\rm tanh}(\chi) \,.
\label{eqn:psi_2_KK}
\end{eqnarray}
Note that $\psi_2 $ is unbounded as $\chi \to \pm \infty $. 
Then using the method of variation of parameters, 
the general solution of  (\ref{eqn:perturb_sys_psi_chi}) can be written in the form
\begin{eqnarray}
\psi  =  \alpha_1 \psi_1 + \alpha_2 \psi_2 -  \frac{\psi_1(\chi)}{W} \int^{\chi }_0 F(\hat{\chi})\psi_2(\hat{\chi})  d\hat{\chi}  
+ \frac{\psi_2(\chi)}{W} \int^{\chi }_0 F(\hat{\chi})\psi_1(\hat{\chi})  d\hat{\chi}  ,
\label{eqn:psi_var_param}
\end{eqnarray}
where $\alpha_{1,2}$ are arbitrary constants.
Since $\psi_1 \sim \exp{(\mp 2\chi )}$ and $\psi_{2} \sim \exp{(\pm 2 \chi )}$ as $\chi \to \pm \infty $,
we must set $\alpha_2 = 0$.  Also the first term in the particular solution will contain a secular term
proportional to $\chi \exp{(\mp 2\chi )}$. This term  originates  from the term in $F$ proportional to $f_0 
\sim \exp{(\mp 2\chi)}$ but can be removed by the choice of $v_1$, the first-order speed correction, namely 
\begin{eqnarray}
v_1  =  \frac{\delta }{\gamma + \delta } [A + (v_{0}^2 -1)C ]\,. 
\label{eqn:kappa}
\end{eqnarray}
Hence we can derive a nonsecular bounded  solution of (\ref{eqn:perturb_sys_psi_chi}): 
\begin{eqnarray}
\psi(\chi)  =  \alpha_1  {\rm sech}^2(\chi)\, {\rm tanh}(\chi) + \frac{3\delta(v_0^2-1)}{\gamma}(A - B  + v_0^2 C){\rm sech}^2(\chi)[1- {\rm tanh}(\chi)].
\label{eqn:psi_gen_soln}
\end{eqnarray}
Importantly this solution is localised in $\chi $, and further by choosing $\alpha_1 = 
3\delta (v_0^2-1)(A - B  + v_0^2 C)/\gamma $ is symmetric in $\chi$. Indeed then it is proportional to $f_0 $
and so by itself simply just determines a first-order correction to the solitary wave amplitude. 
\subsection{The radiating  part of the solution}
\label{sec:soln_phi}
The equation for $\phi$ is more difficult to solve since it is a fourth-order ordinary differential equation
and cannot be reduced to second order by integration.
Let us first write (\ref{eqn:perturb_sys_phi_xi}) in the form
\begin{eqnarray}
\hspace{-8mm} L (\phi ) \equiv \phi_{\xi \xi \xi \xi }  + (3\varepsilon^2\hbox{sech}^2 (\varepsilon \xi /2) \phi)_{\xi \xi } 
 - \varepsilon^2 \phi_{\xi \xi }   - k^4\phi  = Q_{\xi \xi } \,, \ \ \  Q = \frac{P(f_0)}{v^{2}_0 } \,, 
\label{eqn:operator_L}
 \end{eqnarray}
where we recall that  $\varepsilon=\sqrt{v^{2}_0 - 1}/ v_0$ 
and  denote $k^4 = (\delta + \gamma )/v^{2}_0 $.
The homogeneous form of (\ref{eqn:operator_L}) has four linearly independent solutions, which can be uniquely defined
by their behaviour as either  $\xi \to  \infty$ or $\xi \to - \infty$, 
\begin{eqnarray}
\hspace{-10mm}&&\phi_1 \sim \cos{m\xi} \,, \quad \phi_2 \sim \sin{m \xi } \,, \quad \phi_3 \sim \cosh{M\xi } \,,
\quad \phi_4  \sim \sinh{M\xi } \,, \quad \xi \to \infty \, ,
\label{eqn:4sol+}  
\end{eqnarray}
\begin{eqnarray}
\hspace{-10mm}&&\hat{\phi}_1 \sim \cos{m\xi} \,, \quad \hat{\phi}_2 \sim \sin{m \xi } \,, \quad \hat{\phi}_3 \sim \cosh{M\xi } \,,
\quad \hat{\phi}_4 \sim \sinh{M\xi } \,, \quad \xi \to -\infty \,,
\label{eqn:4sol-}
\end{eqnarray}
\begin{eqnarray}
 && \hbox{where} \quad m^2 , M^2 = (k^4 +\frac{\varepsilon^4}{4})^{1/2} \mp \frac{\varepsilon^{2}}{2} \,.
 \label{m}
 \end{eqnarray}
Importantly, note that as $\varepsilon \to 0$,  $m,M \to k$. 
%
Next we exploit the symmetry in the operator $L$, that is,  if $\phi (\xi) $ is a solution,
so too is $\phi (- \xi)$, and so,
$$ \hat{\phi}_1 (\xi ) = \phi_1 (-\xi ) \,, \quad \hat{\phi}_2 (\xi ) =-\phi_2 (-\xi ) \,, \quad 
\hat{\phi}_3 (\xi ) = \phi_3 (-\xi )  \,,  \quad \hat{\phi}_4 (\xi ) = -\phi_4 (-\xi ) \,. $$
Also, it is useful to  note the solutions
\begin{eqnarray}
\phi_5 = \phi_3 - \phi_4 \sim \exp{(-M\xi )},  \quad \xi \to \infty \,; \quad
\phi_6 = \hat{\phi}_3 + \hat{\phi}_4 \sim \ \exp{(M\xi )}, \quad \xi \to \-\infty \,. 
\end{eqnarray}

We again apply the method of variation of parameters, to which end we  express the fourth order equation (\ref{eqn:operator_L}) as a $4 \times 4$ system,
\begin{eqnarray}
 {\bf U }_{\xi } = {\bf A}{\bf U} + {\bf F} \,, \quad {\rm where} \quad {\bf U} = (\phi , \phi_{\xi }, \phi_{\xi \xi }, \phi_{\xi \xi \xi })^{\bf T} \,,
\quad {\bf F} = (0, 0, 0, Q_{\xi \xi})^{\bf T} \ ,
\label{eqn:vector_phi}
\end{eqnarray}
and ${\bf A}$ is a $4 \times 4$ matrix with rows $(0,1,0,0), (0,0,1,0), (0,0,0,1)$,
$((k^4 - S_{\xi \xi }), - 2S_{\xi }, (\varepsilon^2 -S), 0 )$, where $S= 3\varepsilon^2\hbox{sech}^2 (\varepsilon \xi /2)$.
Then we seek a solution of (\ref{eqn:vector_phi}) in the form
\begin{eqnarray}
{\bf U} = B_1 {\bf U}_1 + B_2 {\bf U}_2 + B_3 {\bf U}_3 + B_4 {\bf U}_4 = 
{\bf G}{\bf B} \,, \quad {\bf B} = (B_1 ,B_2 ,B_3 , B_4 )^{\bf T} \,,
\label{eqn:U=GB}
\end{eqnarray}
where the vectors ${\bf U}_i $ for $i=1,2,3,4$ are solutions of the homogeneous part of equation (\ref{eqn:vector_phi}) corresponding to
$\phi_i$, and ${\bf G} $ is the fundamental matrix whose columns are
${\bf U}_1 , {\bf U}_2 , {\bf U}_3 , {\bf U}_{4}$, that is
\begin{eqnarray}
{\bf G}(\xi) = \left(\begin{array}{cccc} 
\phi_1 & \phi_2 & \phi_3 & \phi_4 \\
\phi_{1\xi} & \phi_{2\xi} & \phi_{3\xi} & \phi_{4\xi} \\
\phi_{1\xi\xi} & \phi_{2\xi\xi} & \phi_{3\xi\xi} & \phi_{4\xi\xi} \\
\phi_{1\xi\xi\xi} & \phi_{2\xi\xi\xi} & \phi_{3\xi\xi\xi} & \phi_{4\xi\xi\xi}
\end{array} \right).
\label{eqn:G__eps_limit}
\end{eqnarray}
Substituting (\ref{eqn:U=GB}) into (\ref{eqn:vector_phi}) yields
\begin{eqnarray}
 {\bf B }_{\xi } &=& {\bf G}^{-1} {\bf F} \,,
\label{eqn:B_xi}  
\end{eqnarray}
since ${\bf G}$ (more specifically each column in ${\bf G}$) satisfies the homogeneous equation. Integrating (\ref{eqn:B_xi}) yields  the general solution of (\ref{eqn:vector_phi}):
\begin{eqnarray}
{\bf U} ={\bf G (\xi )}\{{\bf C} + \int^{\xi }_{0} {\bf G}^{-1}(\eta ) {\bf F} (\eta )\, d\eta \} \,, 
\label{eqn:U_sol}
\end{eqnarray}
where ${\bf C} = (C_1, C_2 , C_3 , C_4 )^{\bf T}$ is an arbitrary constant vector. In the sequel, two of these arbitrary constants are chosen to ensure $\phi$ is bounded as $\xi\rightarrow\pm \infty$; that is, they are chosen to remove exponentially  growing terms at infinity. The remaining two constants are associated with oscillatory terms at infinity,
and  these will be chosen to ensure that $\phi \to 0$ as $\xi \rightarrow  \infty$. This is because here the group velocity  $v_g = p +kp_k < p$,} since $p_k < 0$, see Figure \ref{disp_curves_intro}(a), and so radiated waves will appear
only in the region $\xi < 0$, see the numerical solution in Figure \ref{RSW_ch2} and our numerical results  in section 3. Further progress now requires evaluation of ${\bf G}$, that is of $\phi_{i}, i =1,2,3,4$ for all $\xi$. To achieve this we exploit $\varepsilon \ll1 $ in two different ways.

\subsubsection{Approximation of the variable coefficient term} 
\label{sec:RSW_approx_var_term}
Although the fundamental matrix ${\bf G }$ is known when $\xi \to \pm \infty $, 
some specific knowledge is needed around $\xi = 0$ in order to evaluate the integral term in
(\ref{eqn:U_sol}).  The difficulty in finding exact solutions of the homogeneous equation
$L(\phi ) =0 $ is the variable coefficient term  
$3\varepsilon^2 \hbox{sech}^2 (\varepsilon \xi /2) $ on the left-hand side of (\ref{eqn:operator_L}).
In order to make  further analytical progress we make the {\it ad hoc} approximation that 
 this offending term is removed.   Formally we let $\varepsilon \to 0$,
 but only in this  variable coefficient term. This yields approximate solutions of the homogeneous equation in the form
\begin{eqnarray}
\phi_1 \approx  \cos{(m \xi )} \,, \quad \phi_2 \approx \sin{(m \xi )}\,, \quad
\phi_3 \approx \cosh{(M \xi )} \,, \quad \phi_4 \approx \sinh{(M \xi )}\,.
\label{eqn:solns_HE_var_coeff=0}
\end{eqnarray}
This approximation maintains the correct behaviour of $\phi$ as $\xi \to \pm \infty$,  since in the sequel we seek the solution for  $\phi$ only in this far spatial region. 
 Further, with the variable coefficient term neglected in (\ref{eqn:operator_L}), the matrix ${\bf A}$ in equation (\ref{eqn:vector_phi}) becomes  a constant coefficient matrix.
Thus, if ${\bf U}(\xi )$ is a solution of the homogeneous equation, 
 so is ${\bf U}(\xi + \xi_0 )$  for any constant $\xi_0 $. Let 
  ${\bf K}(\xi, \eta) = {\bf G}(\xi ){\bf G}^{-1}(\eta)$ which is 
 the unique matrix solution of the homogeneous equation, 
such that ${\bf K}(\eta , \eta) = {\bf I}$ where ${\bf I}$ is the unit matrix. It then follows that we have 
${\bf K}  (\xi , \eta ) = {\bf E}(\hat{\xi})$, $\hat{\xi} =\xi - \eta $, and ${\bf E}(\hat{\xi })$ is the
unique matrix solution of the homogeneous equation, such that ${\bf E}(0)={\bf I}$.
One can write the unique matrix solution in the form
$${\bf E} (\hat{\xi}) = \frac{1}{m^2 + M^2}({\bf \Phi}_1 , {\bf \Phi}_2, {\bf \Phi}_3, {\bf \Phi}_4 ) \, ,$$
where the column vectors ${\bf \Phi}_i=(\phi_i,\phi_{i\xi},\phi_{i\xi\xi},\phi_{i\xi\xi\xi})^{\bf T}$, for $i=1,2,3,4$, are generated from the approximate solutions (\ref{eqn:solns_HE_var_coeff=0}). 
The first element of each vector is given by
\begin{eqnarray}
\hspace{-6mm}\Phi_1 \hspace{-2mm}&=&\hspace{-2mm} M^2\cos{(m\hat{\xi} )} + m^2\cosh{(M\hat{\xi} )}\,, \quad \Phi_2 = \frac{1}{mM}[M^3\sin {(m\hat{\xi} ) } + m^3\sinh{(M\hat{\xi} )}] \,,   \nonumber \\
\hspace{-6mm} \Phi_3 \hspace{-2mm}&=&\hspace{-2mm} - \cos{(m\hat{\xi} ) } + \cosh{(M\hat{\xi} )}\,, 
\quad \ \ \ \ \   \Phi_4 = \frac{1}{mM}[-M\sin{(m\hat{\xi} ) } + m\sinh{(M\hat{\xi} )}] \,.
\end{eqnarray}
In this approximation, it then follows from (\ref{eqn:U_sol}) that the general solution of (\ref{eqn:vector_phi}) is
\begin{eqnarray}
{\bf U} = {\bf G (\xi )} {\bf C} +  \int^{\xi }_{0} {\bf E}(\hat{\xi}) {\bf F} (\eta )\, d\eta 
= {\bf G (\xi )} {\bf C} + \frac{1}{m^2 + M^2} \int^{\xi }_{0} {\bf \Phi_4}(\hat{\xi})  Q_{\eta\eta} \, d\eta .  
\label{eqn:U_sol_xi_pm_infty}
\end{eqnarray}
Taking the first element  from (\ref{eqn:U_sol_xi_pm_infty}), yields 
an approximate  solution of (\ref{eqn:perturb_sys_phi_xi}),
\begin{eqnarray}
\phi &=& C_1 \cos{(m \xi )} + C_2 \sin{(m \xi )}+ C_3 \cosh{(M \xi )}+ C_4 \sinh{(M \xi )}  \nonumber \\
&+& \frac{1}{m^2 + M^2}\int^{\xi }_{0} \Phi_4 (\hat{\xi })Q_{\eta \eta }(\eta )\, d\eta  \,.
\label{eqn:phi_asym_sol}
\end{eqnarray}
Next the integral term in (\ref{eqn:phi_asym_sol}) can be  integrated twice by parts, 
and noting that $Q$ is a symmetric function of $\eta $, we can write (\ref{eqn:phi_asym_sol}) in the alternative form
\begin{eqnarray}
\phi & = & \tilde{C}_1 \cos{(m \xi )} +  C_2 \sin{(m \xi )}+ \tilde{C}_3 \cosh{(M \xi )}+  C_4 \sinh{(M \xi )}
\nonumber \\
& & + \frac{1}{m^2 + M^2}\int^{\xi }_{0} [m\sin(m\hat{\xi}) + M\sinh(M\hat{\xi})]Q (\eta )\, d\eta  \,,
\label{eqn:phi_asym_sol_alt}
\end{eqnarray}
where $\tilde{C}_1  = C_1 + Q(0)/(m^2+M^2 )$ and  $\tilde{C}_3 = C_3 -  Q(0)/(m^2+M^2 )$.
Then noting that $Q$ is symmetric in $\eta $, and exponentially small in the limit $\xi \rightarrow \pm \infty $,
we get that the  leading order terms at infinity are given by
\begin{eqnarray}
\phi \sim \left(\frac 1 2 \tilde{C}_3 \pm \frac 1 2 C_4 + D\right)\exp{(\pm M\xi )} \,, \quad   {\rm as} \quad  
 \xi \rightarrow \pm \infty ,  
\label{eqn:phi_asym_secular}
\end{eqnarray}
$$
{\rm where} \ \ \ \    D =  \frac{M}{2(m^2+M^2)}\int^{\infty }_{0} \exp(-M\eta )Q(\eta )\,d\eta \,.  
$$
Since these terms are secular, we require $\frac 1 2 \tilde{C}_3 \pm \frac 1 2 C_4 + D=0$, which implies $C_4 =0, \ \tilde{C}_3 = -2D$. Then finally as  $\xi \to \pm \infty $ we find the  oscillatory terms
\begin{eqnarray}
\phi \sim (\tilde{C}_1 + E_3 )\cos{(m\xi )} + (C_2 \pm E_4 )\sin{(m\xi )}\,   \quad  \quad {\rm as} \quad  \quad  \xi \rightarrow \pm \infty , 
\label{eqn:phi_nonsecular_2nd_attempt} 
\end{eqnarray}
 $${\rm where} \ \ \ \ E_3 = -\frac{m}{m^2+M^2}
 \int^{\infty }_{0} \sin{(m\eta )}Q(\eta )\,d\eta \,, \quad 
 E_4 = \frac{m}{m^2+M^2}\int^{\infty }_{0} \cos{(m\eta )}Q(\eta )\,d\eta \,. $$
Finally, we impose an asymmetry condition on $\phi$, namely that we have one-sided radiating  solutions 
in the region $\xi<0$ only. This implies from (\ref{eqn:phi_nonsecular_2nd_attempt}) that 
 $ \tilde{C}_1 = - E_3$ and $C_2=-E_4$, and so
\begin{eqnarray}
\phi &\sim& 0   \qquad \qquad \ \ \ \ \ \ \ \  \quad  {\rm as}  \quad \quad \xi \rightarrow + \infty,   \nonumber  \\
\phi &\sim&  F_{rad1} \sin{(m\xi)}  \quad\quad  {\rm as} \quad  \quad  \xi \rightarrow - \infty.
\label{eqn:aympt_phi_2nd}
\end{eqnarray}
\begin{eqnarray}
F_{rad1} =-2E_4 = - \frac{12m^2\pi [A + Bv_0^2(\varepsilon^2 + m^2) - Cv_0^2m^2]}{(m^2+M^2)\sinh{\left(m\pi /\varepsilon \right)}}\,.
\label{eqn:E2_2nd_attempt}
\end{eqnarray}
This one-sided radiating asymptotic solution for $\phi$ depends only on the original parameters 
in the cRB problem (\ref{eqn:perturb_bous_sys}) without the presence of any arbitrary constants. 
From (\ref{eqn:E2_2nd_attempt})  the amplitude of oscillations  depends on all three perturbations of 
$c^2, \alpha$ and $\beta$ from $1$ and is exponentially small as  $\varepsilon\rightarrow 0$.

\subsubsection{ Asymptotic solution}

\label{sec: asympt_approach_phi}

We next consider an alternative approach to only neglecting the variable coefficient term in the governing equation 
for $\phi$ by systematically exploiting the small parameter $|{\varepsilon}| \ll 1$. 
We first express the equation  (\ref{eqn:operator_L}) in the form
\begin{eqnarray}
\phi''''  - k^4\phi &=&  \varepsilon^2\phi'' (1 -3{\rm sech}^2\chi)  
+6\varepsilon^{3}\phi' {\rm sech}^2\chi {\rm tanh} \chi  + 
\nonumber \\
&&
 \frac{3 \varepsilon^4 }{2}(A-\phi){\rm sech}^2\chi \left(2-3{\rm sech}^2\chi     \right)   + 
 \nonumber \\
 &&\frac{3v_0^2 \varepsilon^6 }{2}{\rm sech}^2\chi  \Big[ 2C +  15C\ {\rm sech}^2\chi ({\rm sech}^2\chi -1)  +
 3B\ {\rm sech}^2\chi (4-5{\rm sech}^2\chi)   \Big] \,,
\label{eqn:phi_eps_no_shift}
\end{eqnarray} 
where $\chi = \frac{\varepsilon}{2} \xi$.
For $\varepsilon=0$ the solution of (\ref{eqn:phi_eps_no_shift}) will exhibit simple harmonic motion 
with some  frequency $\omega$. Hence we use  the Lindstedt--Poinca\'re method 
by introducing a scaled variable $\tau=\omega \xi$ and seeking an asymptotic solution for $\phi(\tau)$, 
while also expanding the frequency $\omega$. 
Under this change of variable in $\xi$ equation  (\ref{eqn:phi_eps_no_shift}) becomes
\begin{eqnarray}
\hspace{-4mm} \omega^4 \phi(\tau)'''' -  k^4\phi(\tau)  &=&   
 \varepsilon^2\omega^2\phi''(\tau) (1 -3{\rm sech}^2\chi)  %
+6\varepsilon^{3}\omega \phi'(\tau) {\rm sech}^2\chi {\rm tanh} \chi   + \nonumber \\
&&   \frac{3\varepsilon^4 }{2}(A-\phi(\tau)) \ {\rm sech}^2\chi \left(2-3{\rm sech}^2\chi     \right)  +
\nonumber \\ 
&&\frac{3v_0^2\varepsilon^6 }{2} {\rm sech}^2\chi \Big[ 2C +
  15C\ {\rm sech}^2\chi ({\rm sech}^2\chi -1)  +3B\ {\rm sech}^2\chi (4-5{\rm sech}^2\chi)   \Big] ,
  ,
\label{eqn:phi_eps_no_shift_tau}
\end{eqnarray} 
where $\chi(\tau)=\varepsilon \tau/ 2\omega $. 
We then  seek an asymptotic solution of (\ref{eqn:phi_eps_no_shift_tau}) as follows
\begin{eqnarray}
\phi(\tau) = \phi_0 +\varepsilon \phi_{1}  +  \varepsilon^2  \phi_2  +  \varepsilon^3  \phi_3   +    \varepsilon^4 \phi_4   +  \varepsilon^5 \phi_5   +  \varepsilon^6 \phi_6   +  \cdots   \ ,
\label{eqn:phi_asym_soln}
\end{eqnarray}
\begin{eqnarray}
\omega = 1 +\varepsilon \omega_{1}  +  \varepsilon^2  \omega_3  +  \varepsilon^3  \omega_3   +    \varepsilon^4 \omega_4   +  \varepsilon^5 \omega_5   +  \varepsilon^6 \omega_6   +  ...   \ .
\label{eqn:phi_asym_soln_freq}
\end{eqnarray}
Substituting (\ref{eqn:phi_asym_soln}) and (\ref{eqn:phi_asym_soln_freq}) into (\ref{eqn:phi_eps_no_shift_tau}) yields 
a sequential set of equations for $\phi_{i}$. Since here we are only concerned with the response 
to the forcing term, which at leading order is $O(\epsilon^4 )$, we can set $\phi_{i} = 0, i = 0,1,2,3$. 
The first relevant equations appear at and beyond $O(\varepsilon^4)$, where we have
\begin{eqnarray}
O(\varepsilon^4): \ \  \phi_4''''  -  k^4 \phi_4 \hspace{-2mm} &=&    \frac{3A}{2} \ {\rm sech}^2\chi \ \left(2-3{\rm sech}^2\chi \right) \,,
\label{eqn:phi4_eqn_shift}
\\
O(\varepsilon^5): \ \   \phi_5''''  -  k^4\phi_5 \hspace{-2mm} &=&  -4\omega_{1}\phi_4'''' \,,
\label{eqn:phi5_eqn_shift}
\\
O(\varepsilon^6): \ \   \phi_6''''  - k^4\phi_6 \hspace{-2mm} &=&\hspace{-2mm}         - (4\omega_{2} + 
6 \omega_{1}^2)\phi_{4}''''    +   \phi_{4}'' (1 -3{\rm sech}^2\chi)  - 4\omega_{1}\phi_{5}''''  \nonumber \\
&+&  \frac{3v_0^2}{2}  {\rm sech}^2\chi \Big[ 2C + 15C {\rm sech}^2\chi ( {\rm sech}^2\chi -1)  
+ 3B\ {\rm sech}^2\chi (4-5{\rm sech}^2\chi)   \Big] .
\label{eqn:phi6_eqn_shift}
\end{eqnarray}
Thus  $\phi_4$ constitutes the leading order solution for $\phi$. It transpires  that non-zero higher order correction terms appear only at subsequent even powers of $\varepsilon$, after imposing boundedness and radiation conditions on the solution at each stage.

\medskip
\noindent
{\bf Leading order solution:}
Again using the method of variation of parameters, the general solution for the leading order term $\phi_4$ is
\begin{eqnarray}
 && \phi_4 =  C_{41} \cos{(k\tau )} + C_{42} \sin{(k\tau )}+ C_{43} \cosh{(k\tau )}+ C_{44} \sinh{(k\tau)}   \nonumber \\
&&  + \frac{1}{2k^3}\Bigg\lbrace  \cos(k\tau) \int^{\tau}_0 L_4(s)\ {\rm sin}(ks) \ ds
- \sin(k\tau)    \int^{\tau}_0 L_4(s)\ {\rm cos}(ks) \ ds    
\nonumber \\
&& +  \frac{e^{k\tau}}{2} \int^{\tau}_0 L_4(s)e^{-ks} \ ds
- \frac{e^{-k\tau}}{2} \int^{\tau}_0 L_4(s)e^{ks} \ ds   \Bigg\rbrace, \,\,
L_4(\tau) = (3A/2) \ {\rm sech}^2\chi  \left(2-3\ {\rm sech}^2\chi \right).
\label{eqn:phi_3_exact}
\end{eqnarray}
Noting that $L_4$ is localised and symmetric, we take the limit $\tau \to \pm\infty$, and get that
\begin{eqnarray}
\phi_4 \sim \left(\frac 1 2 C_{43} \pm \frac 1 2 C_{44} + F_1\right)\exp{(\pm k\tau )}  \,, \quad  {\rm as} \quad  
\tau \rightarrow \pm \infty , \quad  F_1 =  \frac{1}{4k^3}\int^{\infty }_{0} L_4(s)e^{-ks}\,ds \,.
\label{eqn:phi_3_ps}
\end{eqnarray}
Thus for $\phi_4$ to be nonsecular we require $C_{43}=-2F_1$ and $C_{44}=0$. We then find that
\begin{eqnarray}
\phi_4 \sim(C_{41} + F_3 )\cos{(k\tau )} + (C_{42} \pm F_4 )\sin{(k\tau )} \,   \quad  \quad {\rm as} \quad  \quad  \tau \rightarrow \pm \infty , 
\label{eqn:phi_nonsecular_asympt} 
\end{eqnarray}
\begin{eqnarray}
{\rm where} \ \ \ \ F_3 = \frac{1}{2k^3}    \int^{\infty }_{0} L_4(s) \sin{(ks )}\,ds \,, \quad 
 F_4 =  -\frac{1}{2k^3}       \int^{\infty }_{0} L_4(s) \cos{(ks )}\,ds \,, 
 \label{eqn:F3} 
\end{eqnarray}%
Finally we impose  an asymmetric condition on $\phi$, namely that $\phi$ has one-sided oscillations only in the region $\xi<0$. This requires that $C_{41}= - F_3$ and $C_{42}=-F_4$. 
This reduces  (\ref{eqn:phi_nonsecular_asympt}) to
\begin{eqnarray}
\phi_4 &\sim& 0   \qquad  \ \ \ \ \ \ \  \ \ \ \ \   \ \quad  {\rm as}  \quad \quad \tau \rightarrow + \infty \,,   \nonumber  \\
\phi_4 &\sim&   F_{rad} \sin{(k\tau)} \quad\quad  {\rm as} \quad  \quad  \tau \rightarrow - \infty \,, 
\label{eqn:aympt_phi_asympt_approach}
\end{eqnarray}
where from (\ref{eqn:F3}), 
\begin{eqnarray}
 F_{rad} = - 2 F_4&=& - \frac{6A\pi \omega^4 }{\varepsilon^4 \  {\rm sinh}\left(k \omega \pi /\varepsilon \right)}\,.
\label{eqn:F4}
\end{eqnarray}
In the limit $\varepsilon \to 0$ this  leading order asymptotic solution for 
$\phi \sim \varepsilon^4 \phi_4 $  coincides with the previous solution (\ref{eqn:aympt_phi_2nd})  found from the alternative approach in section \ref{sec:RSW_approx_var_term} only when $B=C$.
Indeed, in this present approach, any dependence on $B, C$ can only be found at higher order.

\medskip
\noindent
{\bf Higher order correction terms}:  
It is clear that the terms on the right-hand side of the higher order equations (\ref{eqn:phi4_eqn_shift}) and (\ref{eqn:phi5_eqn_shift}), which do not tend to zero in the limit $\tau \to \pm \infty$, will 
produce  secular terms. Using the same approach as for the leading order equation, we find in order to impose an asymmetric condition, some of these secular terms are non-removable by any choice of arbitrary constants alone,
where  these arbitrary constants arise from the solutions of the homogeneous equations. 
Instead we need to use  the higher order correction terms in the expansion of the frequency $\omega$.

At $O(\varepsilon^5)$ we can easily remove these  secular terms by choosing $\omega_{1}=0$, and as a result, (\ref{eqn:phi4_eqn_shift}) reduces to a homogeneous equation  in the same form as that for $\phi_{i}, i=0,1,2,3$ 
which implies that $\phi_5 = 0$. 
With this choice of $\omega_{1}$, the equation (\ref{eqn:phi6_eqn_shift}) at $O(\varepsilon^6)$, reduces to
\begin{eqnarray}
\phi_6''''  - k^4\phi_6   
&=&   -3 \phi_4''\ {\rm sech}^2\chi    +   P(\tau)  +  L_{6BC}(\tau),
\quad P(\tau)=\phi_4''  -  4\omega_2\phi_4'''' \,,
\label{eqn:aympt_phi_6_tau}
\end{eqnarray}
\begin{eqnarray}
\hspace{-2mm}L_{6BC}(\tau) = \frac{3v_0^2}{2}  {\rm sech}^2\chi \left[ 2C + 15C {\rm sech}^2\chi ( {\rm sech}^2\chi -1)  +3B\ {\rm sech}^2\chi (4-5\ {\rm sech}^2\chi)   \right] \hspace{-1mm}.
\label{eqn:L6BC}
\end{eqnarray}
Note that $L_{6BC}$ depends only on $B,C$ while the dependence on $A$ is contained in $P$.  The 
undetermined $\omega_2 $ occurs only in $P(\tau )$. Here in $P(\tau )$ as $\tau \to \pm \infty $
there are trigonometric terms proportional to ${\rm cos}(k\tau), {\rm sin}(k\tau )$ and 
these will produce secular terms in $\phi_6 $ unless they are removed. This can be achieved 
by choosing $\omega_2 = -1/4k^2 $ so that $\omega = 1 - \varepsilon^2 /4k^2 + \cdots $.
Then $k\tau = k\omega \xi = k( 1 - \varepsilon^2 /4k^2 + \cdots )\xi $ agrees with
the exact expression $m\xi $ when $m $ is expanded in powers of $\varepsilon^2 $, 
see (\ref{m}).

With $\omega_1, \omega_2 $ now determined, the general solution of 
({\ref{eqn:aympt_phi_6_tau}) can be found and secular terms removed as for the
leading order term. We omit the extensive details, and again imposing the asymmetric 
condition of one-sided oscillations, we find that
\begin{eqnarray}
\phi_6 & \sim &0   \quad  \qquad\qquad\qquad\qquad\qquad\,\,  \ {\rm as} \quad \tau \rightarrow + \infty,   \nonumber \\
\phi_6 & \sim & -2K_{1}\cos(k\tau)   - 2 K_{2}\sin(k\tau)   \quad  \quad \hspace{-1mm}  {\rm as}  \quad  \tau \rightarrow - \infty.
\label{eqn:phi_6_xi->+infty_osc_asym}
\end{eqnarray}
Here $K_1, K_2 $ are constants given by Fourier-type integrals where the integrand is either 
$P(\tau)$ or $L_{6BC} (\tau)$. Since $P(\tau)$ is proportional to $A$, while only $L_{6BC}$ contains $B, C$
which is the focus of interest here, we need to find only  those parts of $K_1, K_2 $ which arise from
$L_{6BC}$. It turns out that there is no such dependence in $K_1 $, while the contribution to $K_2 $ is
\begin{eqnarray}
\hspace{-2mm}K_{2BC}  \hspace{-2mm}&=&\hspace{-2mm}  -\frac{1}{2k^3}    
\int^{\infty}_{0} \hspace{-2mm}  \cos(ks) L_{6BC}(s)ds  = 
\frac{3v_0^2 \pi[ B(\varepsilon^2 + k^2) - Ck^2  ]   }{  \varepsilon^6 \,{\rm sinh}\left(\frac{ k\omega \pi }{ \varepsilon }\right)  }.
\label{eqn:L6BC_int_exact}
\end{eqnarray}
Thus the asymptotic solution of (\ref{eqn:phi_eps_no_shift_tau}) as $\tau \to \pm \infty$, 
up to $O(\varepsilon^6)$, and keeping only the contribution of $K_{2BC}$ to $K_2$ yields
\begin{eqnarray}
\hspace{-4mm}\phi & \sim & 0   \qquad  \qquad \qquad \quad 
{\rm as} \quad \xi \rightarrow + \infty,   \nonumber \\
\hspace{-4mm}\phi & \sim & F_{rad2} \sin(k \omega \xi)   
\quad  \quad  \,\,  {\rm as}  \quad  \xi \rightarrow - \infty.
\label{eqn:final_asym_nonshift}
\end{eqnarray}
\begin{eqnarray}
{\rm where } \quad  F_{rad2} =  -2\varepsilon^4F_4   -   2\varepsilon^6   K_{2BC} =
-\frac{6 \pi[ A\omega^4 + B v_{0}^2 (\varepsilon^2 + k^2) - C v_{0}^2 k^2  ]   }
{\,{\rm sinh}\left(\frac{ k\omega  \pi }{ \varepsilon }\right)  }
\label{coeff-final}
\end{eqnarray}
Further corrections to the asymptotic solution for $\phi$ and $\omega $ can be obtained by 
systematically carrying the expansion beyond $O(\varepsilon^6)$. 
Note that although $\phi $ is $O(\varepsilon^4 )$ in the localised central core, 
the radiating tail amplitude $F_{rad2}$ has no such dependence on the power $\varepsilon^4 $
but is exponentially small as $\varepsilon \to 0$. 
Further this amplitude agrees with that obtained by the previous method in section 2.2.1
as $\varepsilon \to 0$; see (\ref{eqn:aympt_phi_2nd}), (\ref{eqn:E2_2nd_attempt}) and note that 
as $\varepsilon \to 0$, $m, M \to k$ and $\omega \to 1$.

\section{Numerical simulations of radiating solitary waves}

Returning to the original variables $f$ and $g$, we have the following asymptotic solution of the system of cRB equations (\ref{fg}),
\begin{eqnarray}
f \hspace{-1mm}=\hspace{-1mm} A_0\  {\rm sech}^2 (\chi)+\frac{\mu}{\delta + \gamma}(\gamma \psi + \delta \phi) + O(\mu^2), \ \ \ 
g \hspace{-1mm}=\hspace{-1mm} A_0\  {\rm sech}^2(\chi) + \frac{\mu\gamma}{\delta + \gamma}(\psi - \phi) + O(\mu^2)  .  \nonumber
\label{eqn:perturb_exp_f0}
\end{eqnarray} 
Here we use the localised solution  (\ref{eqn:psi_gen_soln}) for $\psi $, and then either the 
asymptotic radiating solution for $\phi $ obtained in section \ref{sec:RSW_approx_var_term}, given by (\ref{eqn:aympt_phi_2nd}), or the analogous expression obtained in section \ref{sec: asympt_approach_phi} given by (\ref{eqn:final_asym_nonshift}).

In this section we compare these  theoretically derived radiating solitary wave solutions of the cRB equations (\ref{fg}) with corresponding numerical simulations, using a pseudo-spectral method which is extended from the work in \citet{Engelbrecht2} for a single regularised Boussinesq equation.
We let $x\in [-L,L]$, for finite $L$, and discretise the $(x,t)$ domain into a grid with constant spacings $\Delta x$ and $\Delta t$. The solutions $f(x,t)$ and $g(x,t)$ of the cRB equations (\ref{fg}) are approximated by the solutions $f(i\Delta x,j\Delta t)=f_{\rm num}$ and $g(i\Delta x,j\Delta t)=g_{\rm num}$ for $i=1,2,...,N$ and $j=0,1,...$, found 
from  the spectral method.
Since we use  a spectral method, we impose periodic boundary conditions, but in order to simulate the solutions propagating on the infinite line, we choose $L$ to be sufficiently large. To compare the numerical and theoretical solutions we choose the initial conditions in numerical simulations to coincide with the localised part 
of the theoretical solutions (\ref{eqn:perturb_exp_f0}).  
This comprises of the leading order solitary wave solution and the higher order correction terms given by $\psi$, 
obtained in section \ref{sec:RSW_non_osc_phi}.

\subsection{Simulations compared with  the theory in section \ref{sec:RSW_approx_var_term}}

We first consider the radiating solitary wave solutions derived from the first approach, which we denote as 
$f_{\rm theory}$, $g_{\rm theory}$. Figures \ref{figure:RSW_perturb_c} and \ref{figure:RSW_perturb_beta} depict 
a typical comparison of the theoretical and numerical radiating solitary wave solutions propagating in each component $f$ and $g$.  
In both figures we find the leading order solitary wave solution is indeed significantly improved by the localised higher order correction term $\psi$. However, as one can see from the close up plots of the oscillating tail region, there are significant discrepancies in the amplitude of oscillations, most  evident in Figure \ref{figure:RSW_perturb_c}, which corresponds to perturbations in $c$. 
On the other hand  the wavenumber of the oscillations from the theoretical ($m$) and numerical solution are in good agreement, as well as with predictions found from the dispersion relation (see Table \ref{table:RSW_perturb_c_wavenumbers}). 
Although the tail region of the radiating solitary waves shown in Figures \ref{figure:RSW_perturb_c} and  \ref{figure:RSW_perturb_beta} are rather small, they are still considerably greater in magnitude than $O(\mu^2)$, which rules out that the discrepancies are due to higher order terms in the 
expansions for $f$ and $g$, see (\ref{eqn:perturb_exp}), not being included.
\begin{table} [htbp]
\centering
\begin{tabular}{ c  c  c } \hline  \vspace{-4.5mm} \\ 
 & Wavenumber  & Wavelength \\ \hline
Dispersion Relation&  0.8110  &   7.747   \\ 
Numerical &  0.81 $\pm$ 0.1  &    7.75 $\pm$ 0.1   \\
$k$ & 0.85 &     7.392     \\
$m$ & 0.8147 &   7.712   \\
$k\omega$ 
& 0.8139 &   7.719   \\  \hline
\end{tabular}
\caption{\small Comparison of wavenumbers and wavelengths of the numerical solution (averaged readings), $m$, $k$ and that predicted from the linear dispersion relation. All parameter values correspond to Figure \ref{figure:RSW_perturb_c}.}
\label{table:RSW_perturb_c_wavenumbers}
\end{table}
\begin{figure}[htbp]
\begin{center}$
\begin{array}{ccc}
\includegraphics[width=2.4in]{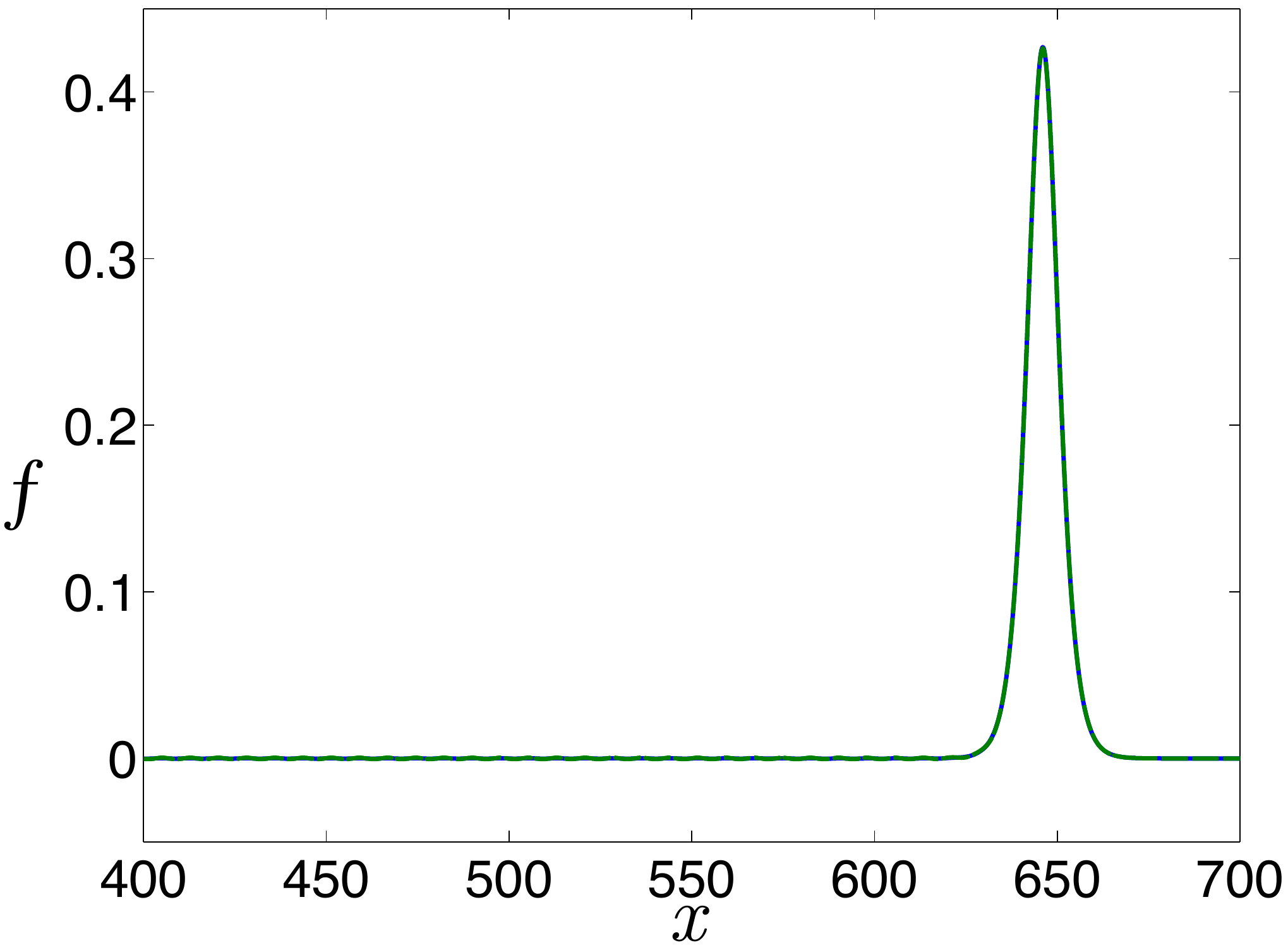}  &
\includegraphics[width=2.35in]{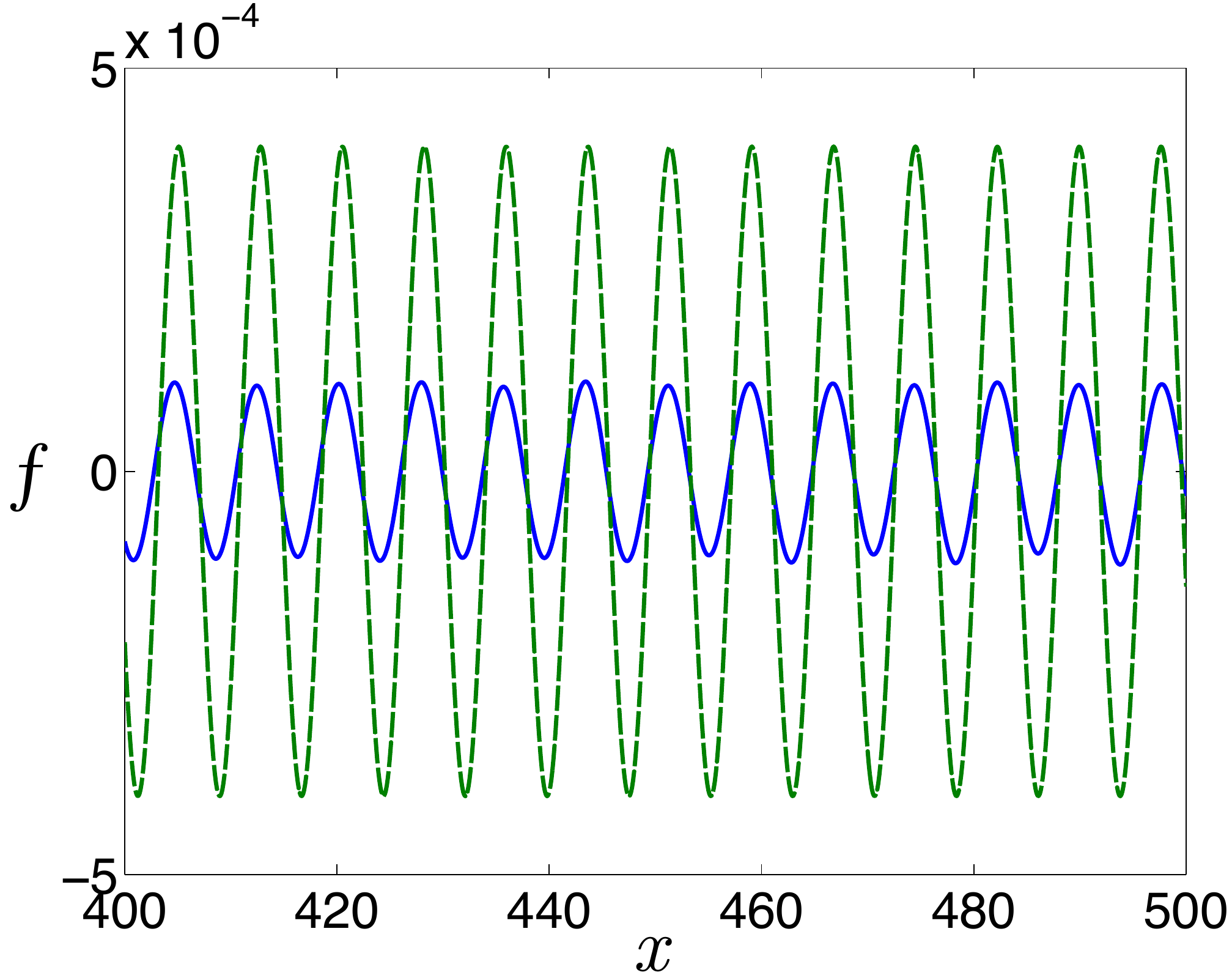}  \\
 \mbox{(a) \ \footnotesize \bf $f_{{\rm num}}$({\color{blue} \bf ---}) \  $f_{{\rm theory}}$({\color{KMgreen} \bf - -}) }   
 & \mbox{(b) \  \footnotesize \bf $f_{{\rm num}}$({\color{blue} \bf ---}) \  $f_{{\rm theory}}$({\color{KMgreen} \bf - -}) }
\\
\includegraphics[width=2.4in]{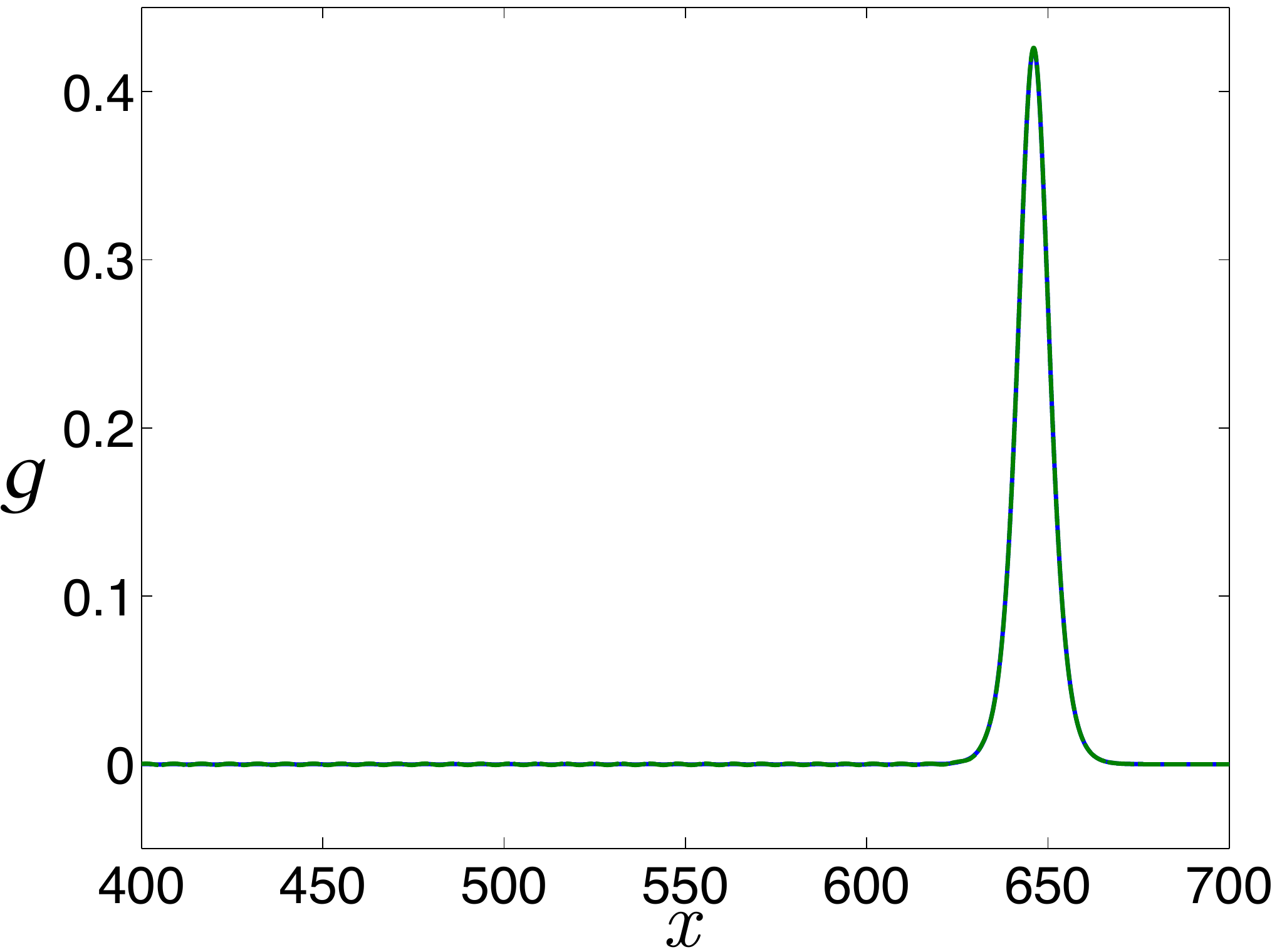}  &
\includegraphics[width=2.35in]{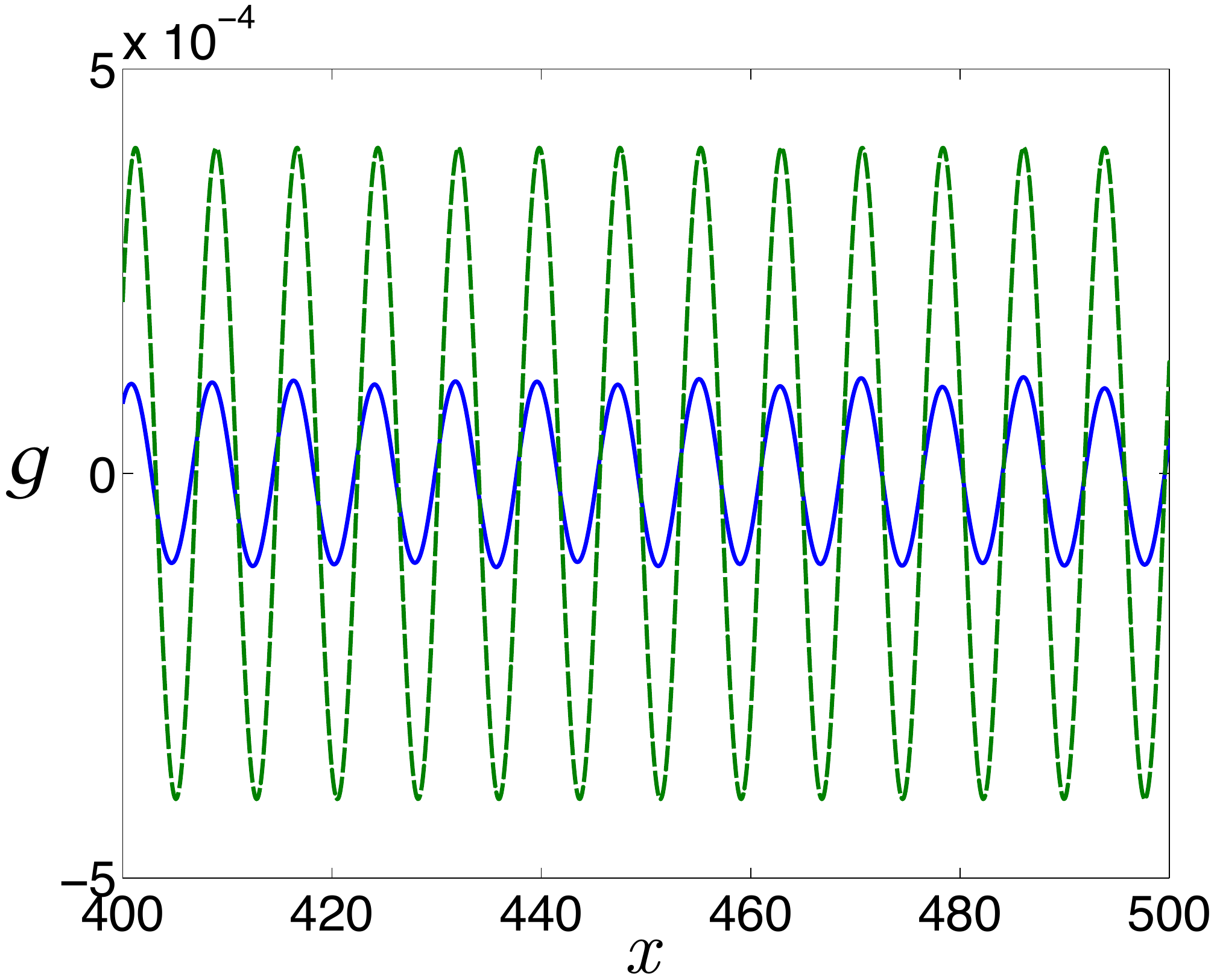}  \\
 \mbox{(c) \ \footnotesize \bf $g_{{\rm num}}$({\color{blue} \bf ---}) \  $g_{{\rm theory}}$({\color{KMgreen} \bf - -}) }   
 & \mbox{(d) \  \footnotesize \bf $g_{{\rm num}}$({\color{blue} \bf ---}) \  $g_{{\rm theory}}$({\color{KMgreen} \bf - -}) }
\end{array}$
\end{center}
\caption{\small Numerical solution and theoretical solutions at  $t=600$ and a magnification of the oscillating tail. Parameter values: $\varepsilon=0.35$, $k = 0.85$, $\mu=0.005$ and $A=7$, $B=C=0$, which implies $m = 0.8147$, $M = 0.8868$, $\omega=0.9576$, $c = 1.017$, $\alpha=\beta=1$, $\gamma=\delta = 0.297$, $v = 1.077$, $v_0 = 1.068$. Numerical parameters: $\Delta t = 0.01$, $L=2000$, $N=4\times10^5$.}
\label{figure:RSW_perturb_c}
\end{figure}
\begin{figure}[htbp]
\begin{center}$
\begin{array}{ccc}
\includegraphics[width=2.4in]{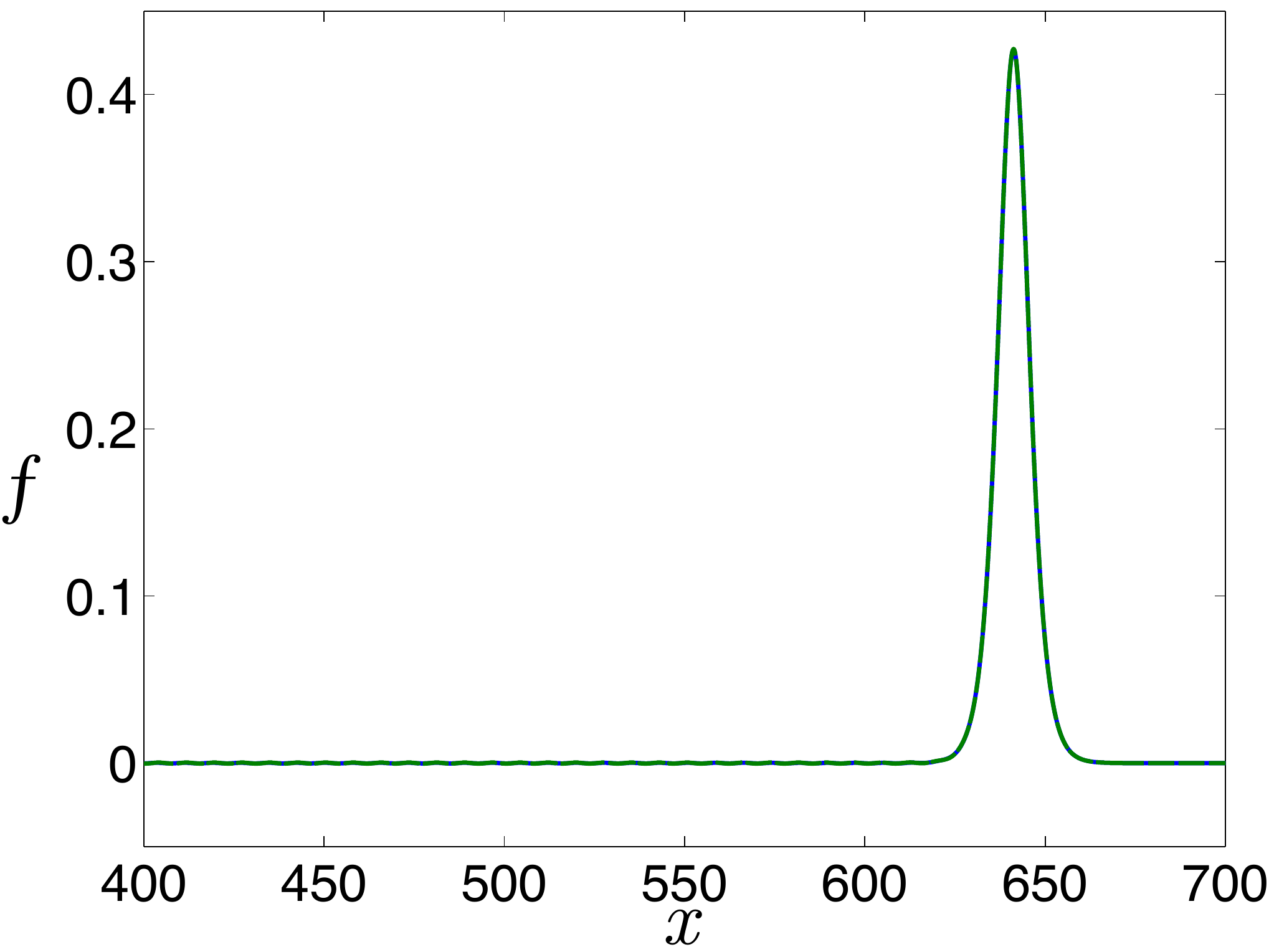}  &
\includegraphics[width=2.35in]{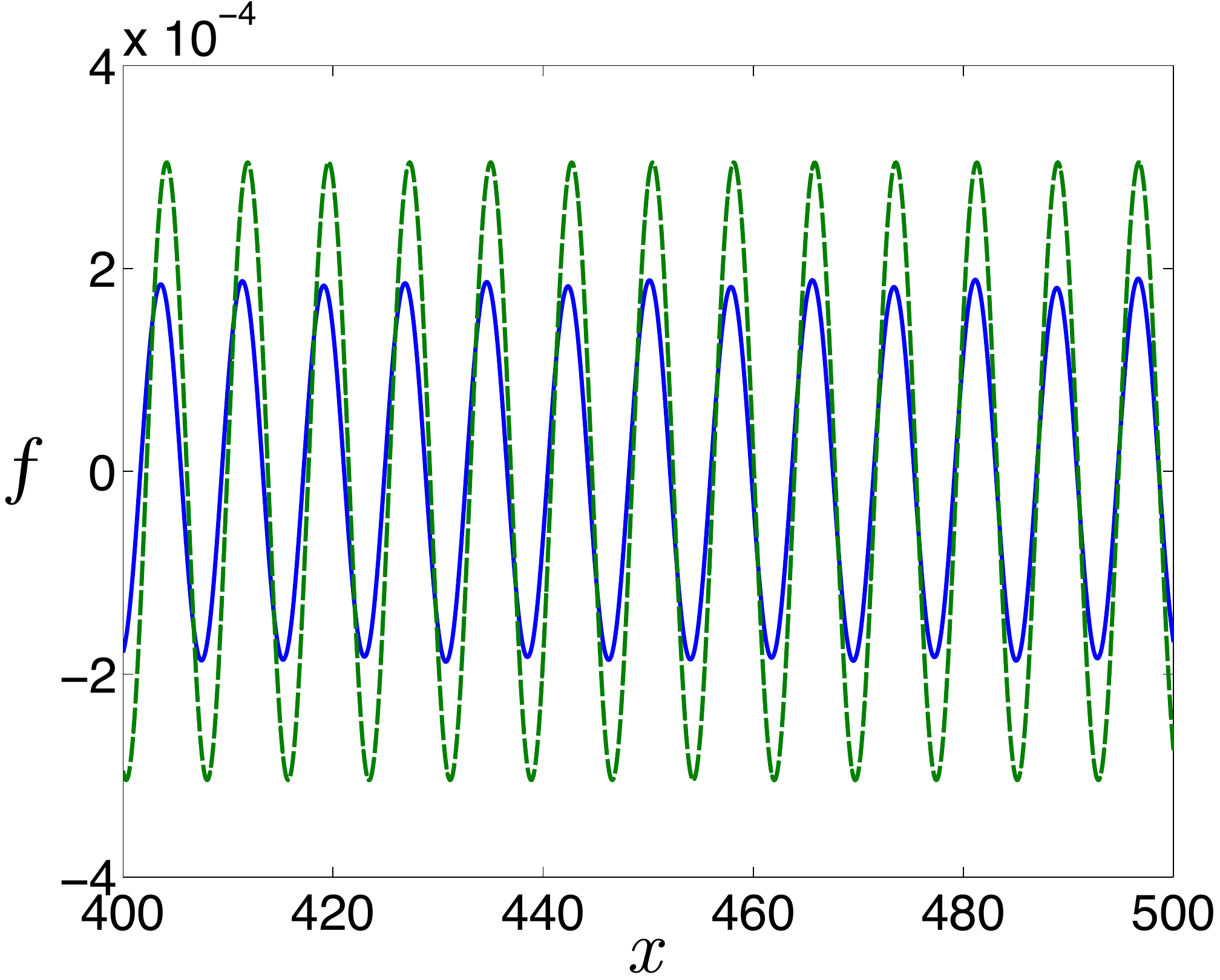}  \\
 \mbox{(a) \ \footnotesize \bf $f_{{\rm num}}$({\color{blue} \bf ---}) \  $f_{{\rm theory}}$({\color{KMgreen} \bf - -}) }   
 & \mbox{(b) \  \footnotesize \bf $f_{{\rm num}}$({\color{blue} \bf ---}) \  $f_{{\rm theory}}$({\color{KMgreen} \bf - -}) }
\\
\includegraphics[width=2.4in]{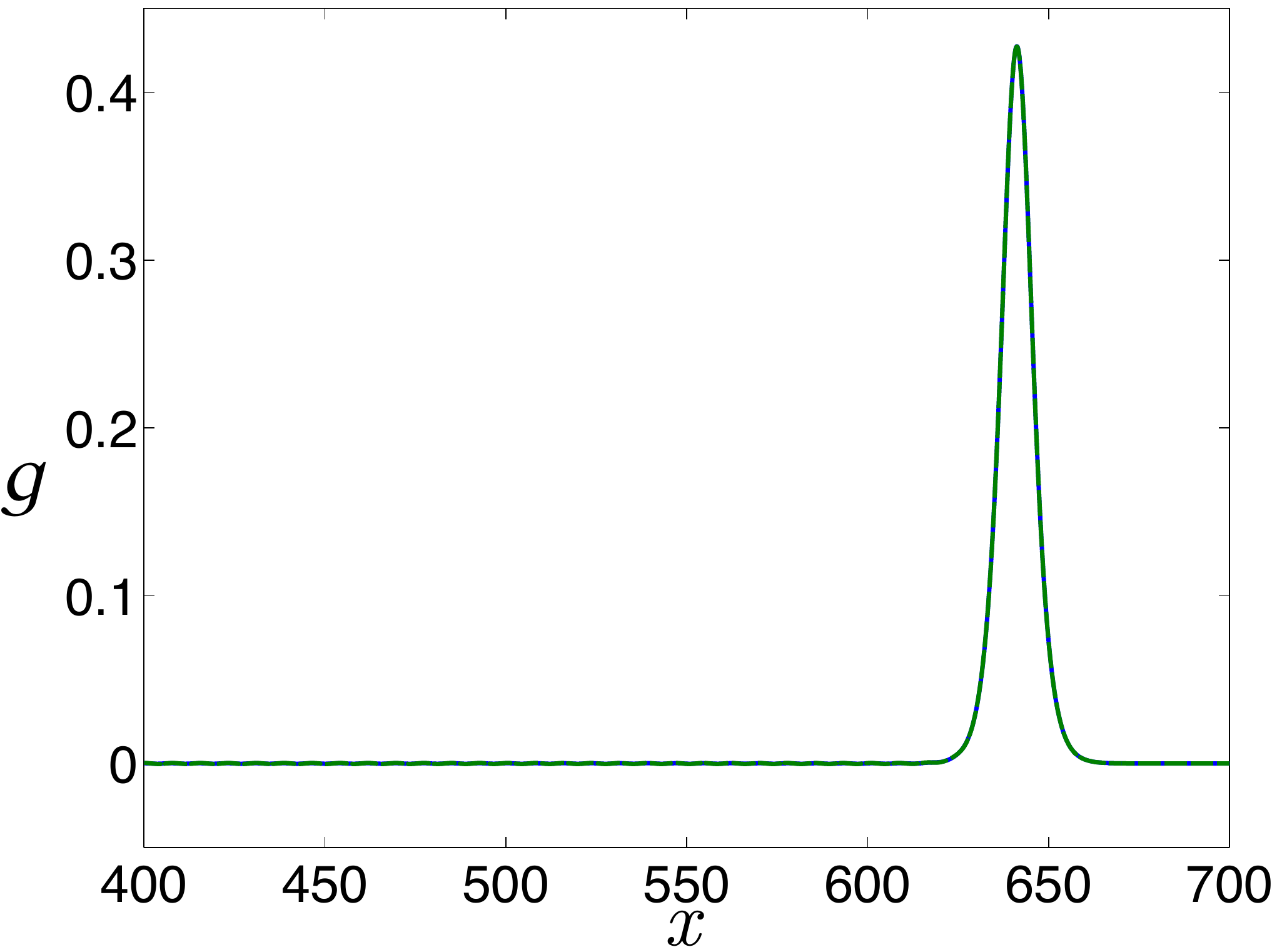}  &
\includegraphics[width=2.35in]{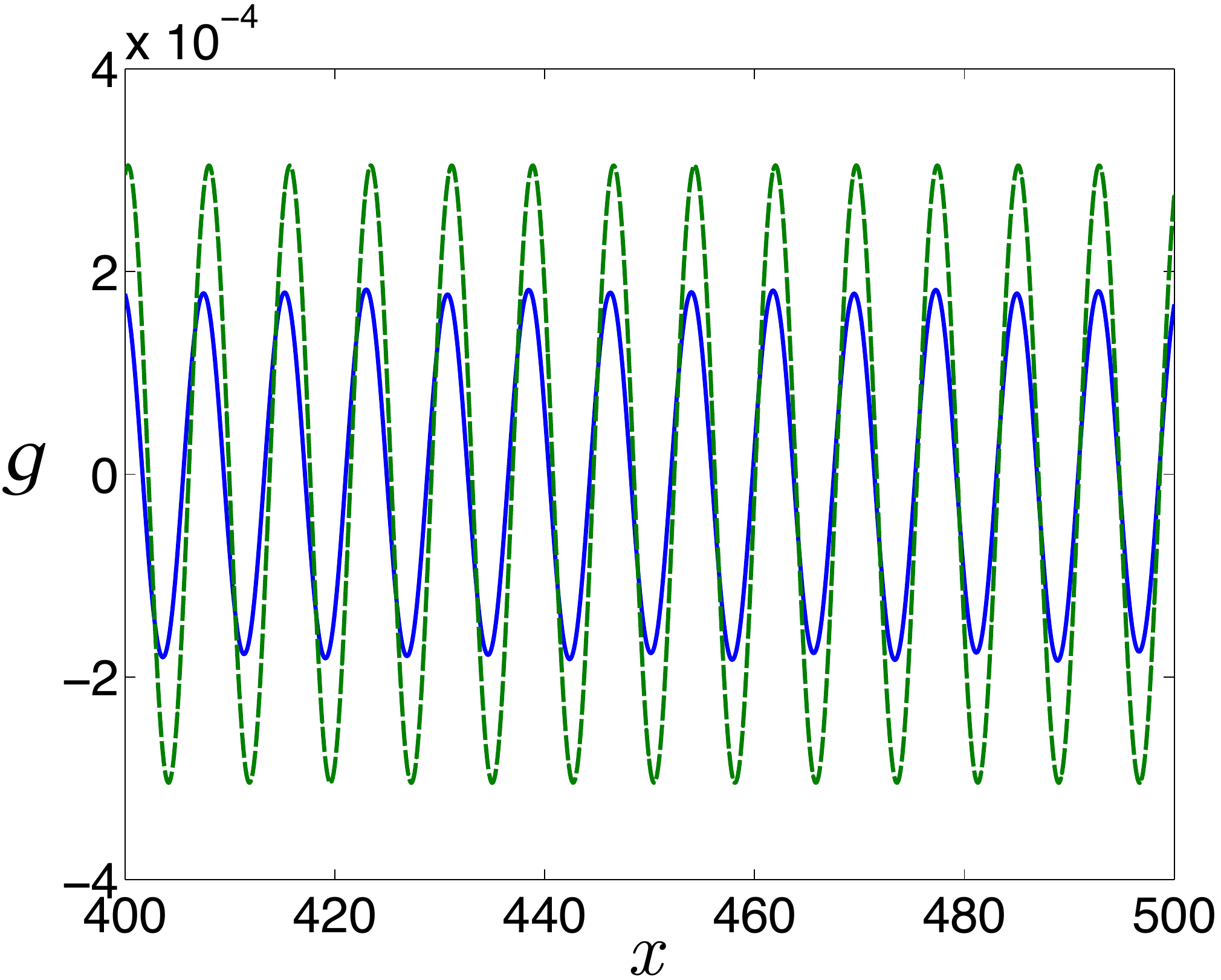}  \\
 \mbox{(c) \ \footnotesize \bf $g_{{\rm num}}$({\color{blue} \bf ---}) \  $g_{{\rm theory}}$({\color{KMgreen} \bf - -}) }   
 & \mbox{(d) \  \footnotesize \bf $g_{{\rm num}}$({\color{blue} \bf ---}) \  $g_{{\rm theory}}$({\color{KMgreen} \bf - -}) }
\end{array}$
\end{center}
\caption{\small Numerical solution and theoretical solution  at  $t=600$ and a magnification of the oscillating tail. Parameter values: $\varepsilon=0.35$, $k = 0.85$, $\mu=0.005$ and $C=7$, $A=B=0$, which implies $m = 0.8147$, $M = 0.8868$, $\omega=0.9576$, $c = \alpha=1$, $\beta=1.035$, $\gamma=\delta = 0.297$, $v = 1.069$, $v_0 = 1.068$. Numerical parameters: $\Delta t = 0.01$, $L=2000$, $N=4\times10^5$.}
\label{figure:RSW_perturb_beta}
\end{figure}

\subsection{Simulations compared with  the theory in  section \ref{sec: asympt_approach_phi}}

Next we consider the alternative asymptotic approach for finding $\phi$ where the outcome is given by (\ref{eqn:final_asym_nonshift}). 
Figure \ref{figure:RSW_perturb_2nd} depicts typical numerical simulations and the theoretical solution for the same perturbation in the parameter $c$ as in the simulations illustrated in Figure \ref{figure:RSW_perturb_c}, which is reincluded in Figure \ref{figure:RSW_perturb_2nd} for comparison purposes. We denote the 
asymptotic solutions $f$ and $g$ obtained from this  perturbation solution for $\phi$ as 
$f^i_{\rm pert}$ and $g^i_{\rm pert}$, where $\phi$ is taken up to $O(\varepsilon^{i})$, for $i=4,6$.

One can see that the solutions for $f$ and $g$ with the higher order terms for $\phi$ included indeed correct the leading order approximations $f^1_{\rm pert}$ and $g^1_{\rm pert}$. This supports the approach outlined in section \ref{sec: asympt_approach_phi} as a valid perturbation method for obtaining the oscillatory part of the radiating solitary wave solutions. Moreover, it is clear that this alternative asymptotic approach for finding the oscillatory part of radiating solitary wave solutions of cRB equations is more effective. The solutions for $f$ and $g$ with $\phi$ taken up to $O(\varepsilon^6)$ significantly improves the discrepancy in the amplitude of oscillations from the numerical
solution, compared with the solutions for $f$ and $g$ using $\phi$ from the previous approach in section \ref{sec:RSW_approx_var_term}.

Further, we also see an improvement in the amplitude of oscillations in the solutions 
$f^1_{\rm pert}$ and $g^1_{\rm pert}$, compared with the solutions using 
$\phi$ from the previous approach denoted here as $f_{\rm theory}$ and $g_{\rm theory}$. 
It was previously noted that the leading order perturbation solution in $\phi$ is equivalent to the solution for $\phi$ from the previous approach when $\varepsilon \to 0$.  However this is only for leading order expressions 
$\omega \approx 1$, $m, M \approx k$. 
Since $\omega$ appears explicitly in $F_{rad2}$ (\ref{coeff-final}), taking higher order terms in 
$\phi$ and  $\omega$, will improve the amplitude of oscillations even for the leading order solution. 
This is evident in Figure \ref{figure:RSW_perturb_2nd}, which displays the solutions $f^{1,2}_{\rm pert}$, 
$g^{1,2}_{\rm pert}$ with the asymptotic solution $\phi$ 
found with the first non-zero correction term in $\omega$ (and with $m$ and $M$ taken exactly 
as in the first approach for finding $\phi$), where one can already notice a distinct improvement in the accuracy of the leading order solutions $f^{1}_{\rm pert}$, $g^{1}_{\rm pert}$ from the solutions using the previous 
approach in finding $\phi$.  

The same follows for perturbations in the parameter $\beta$ as depicted in Figure \ref{figure:RSW_perturb_beta_2nd}. In this case $A=0$, and so the leading order asymptotic solution for $\phi$ appears at $O(\varepsilon^6)$, thus the solutions denoted $f^2_{\rm pert}$, $g^2_{\rm pert}$ correspond to the leading order approximations for $\phi$ and are already relatively close to the numerical solution. 

\begin{figure}[htbp]
\begin{center}$
\begin{array}{ccc}
\includegraphics[width=2.4in]{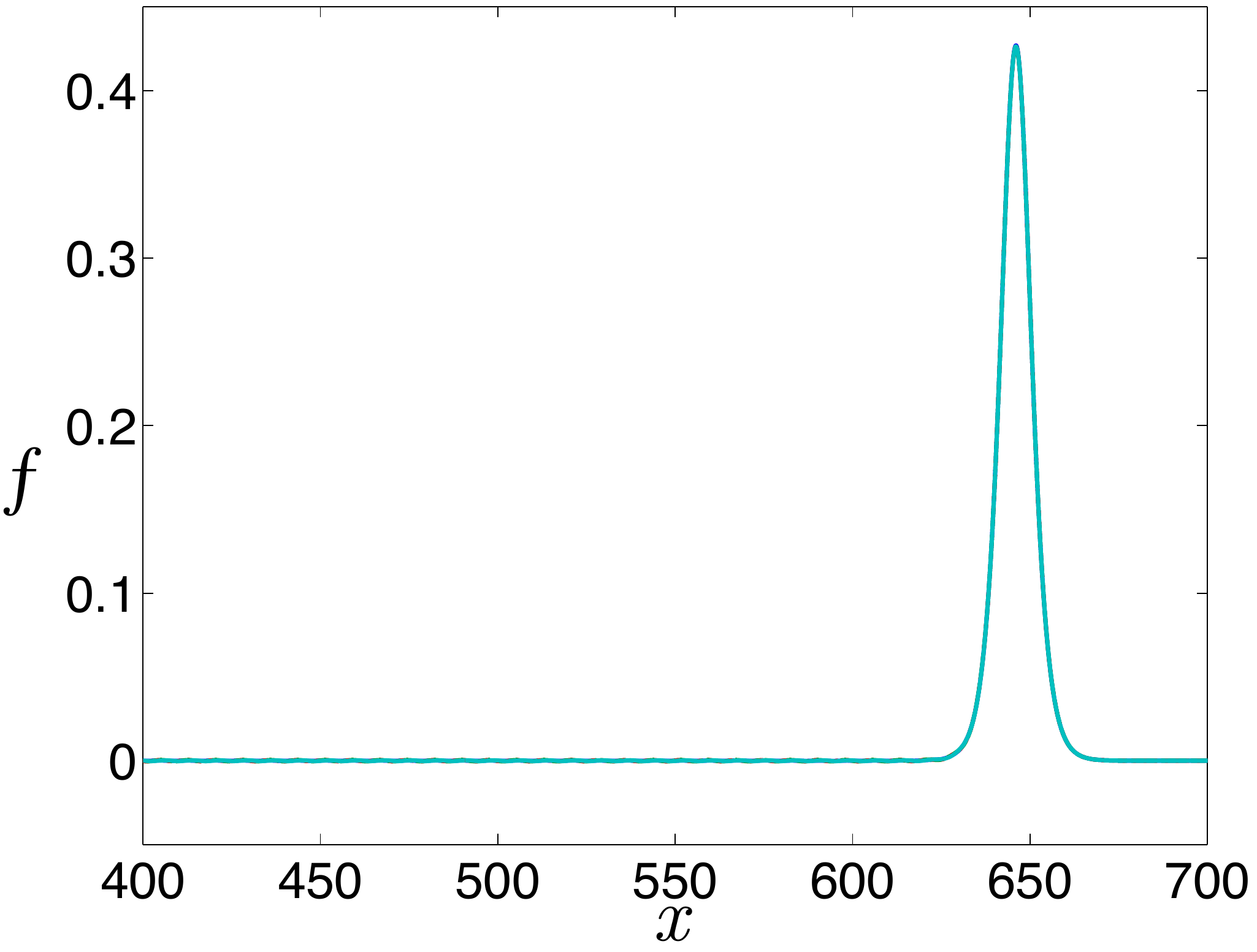}  &
\includegraphics[width=2.35in]{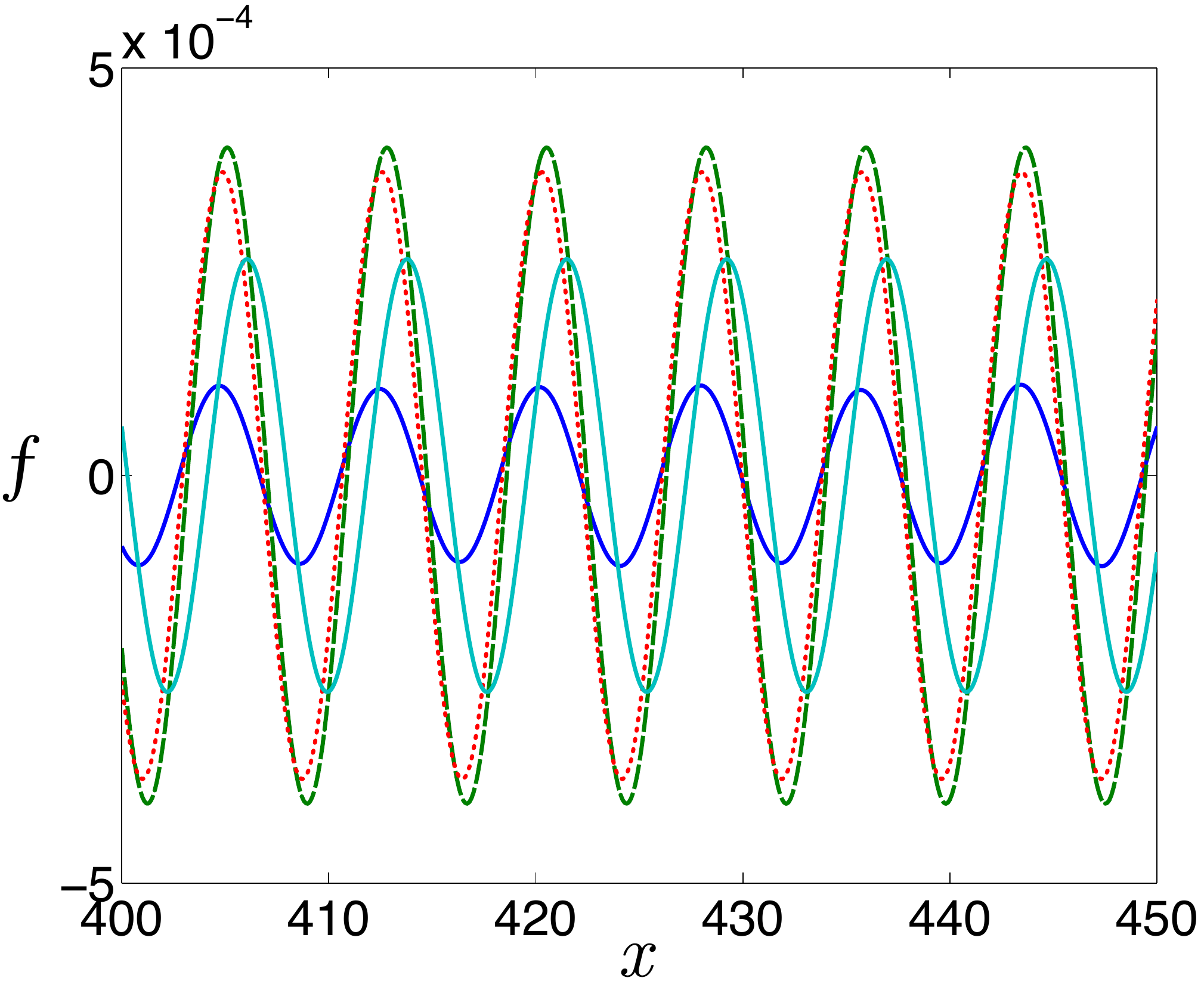}  \\
 \mbox{(a) \ \footnotesize \bf $f_{{\rm num}}$({\color{blue} \bf ---}) \  $f_{{\rm theory}}$({\color{KMgreen} \bf - -}) \ $f^1_{{\rm pert}}$$({\color{red} \bf \cdots})$ \  $f^2_{{\rm pert}}$({\color{Turquoise} \bf ---}) }   
 &  \mbox{(b) \ \footnotesize \bf $f_{{\rm num}}$({\color{blue} \bf ---}) \  $f_{{\rm theory}}$({\color{KMgreen} \bf - -}) \ $f^1_{{\rm pert}}$$({\color{red} \bf \cdots})$ \  $f^2_{{\rm pert}}$({\color{Turquoise} \bf ---}) }  
\\
\includegraphics[width=2.4in]{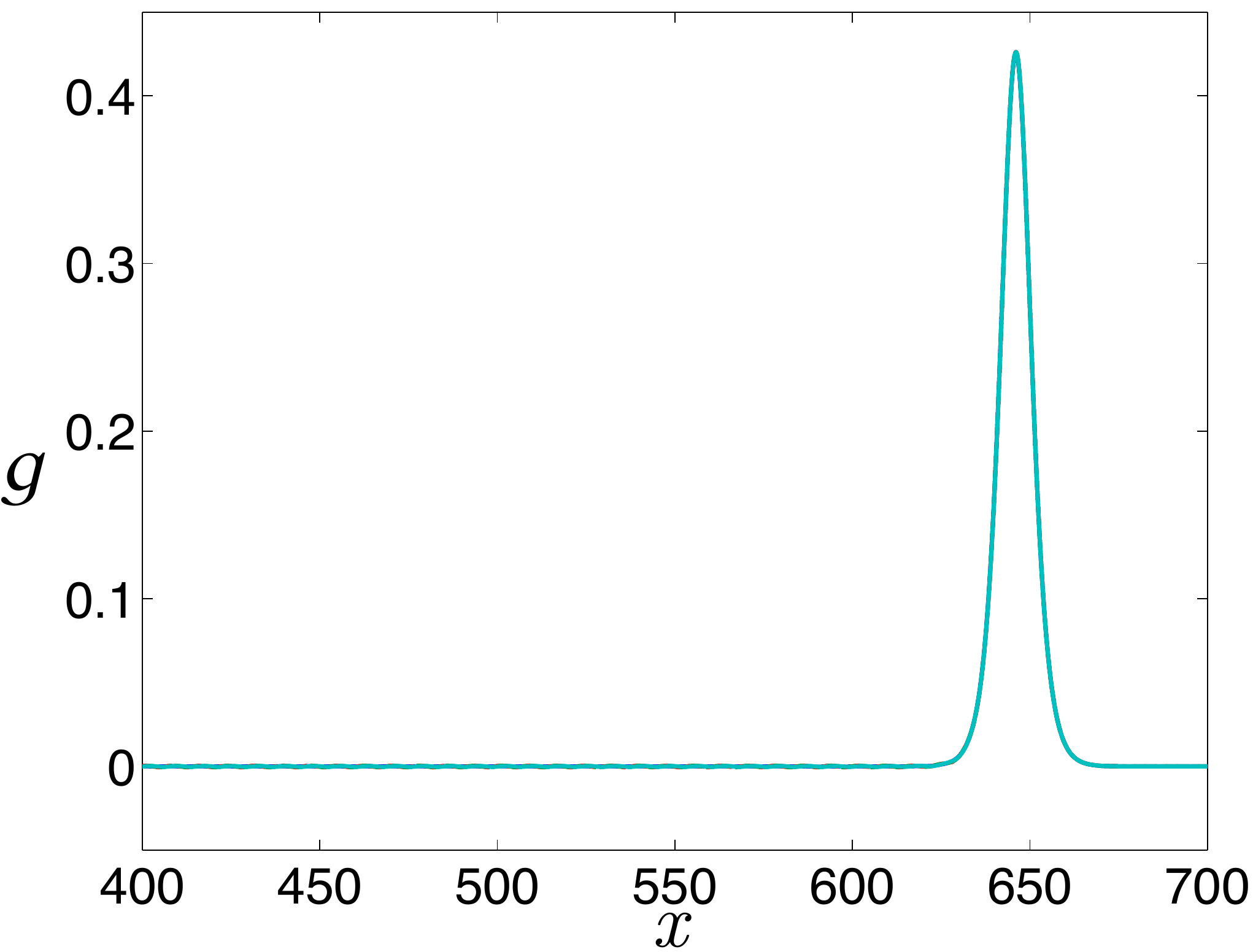}  &
\includegraphics[width=2.35in]{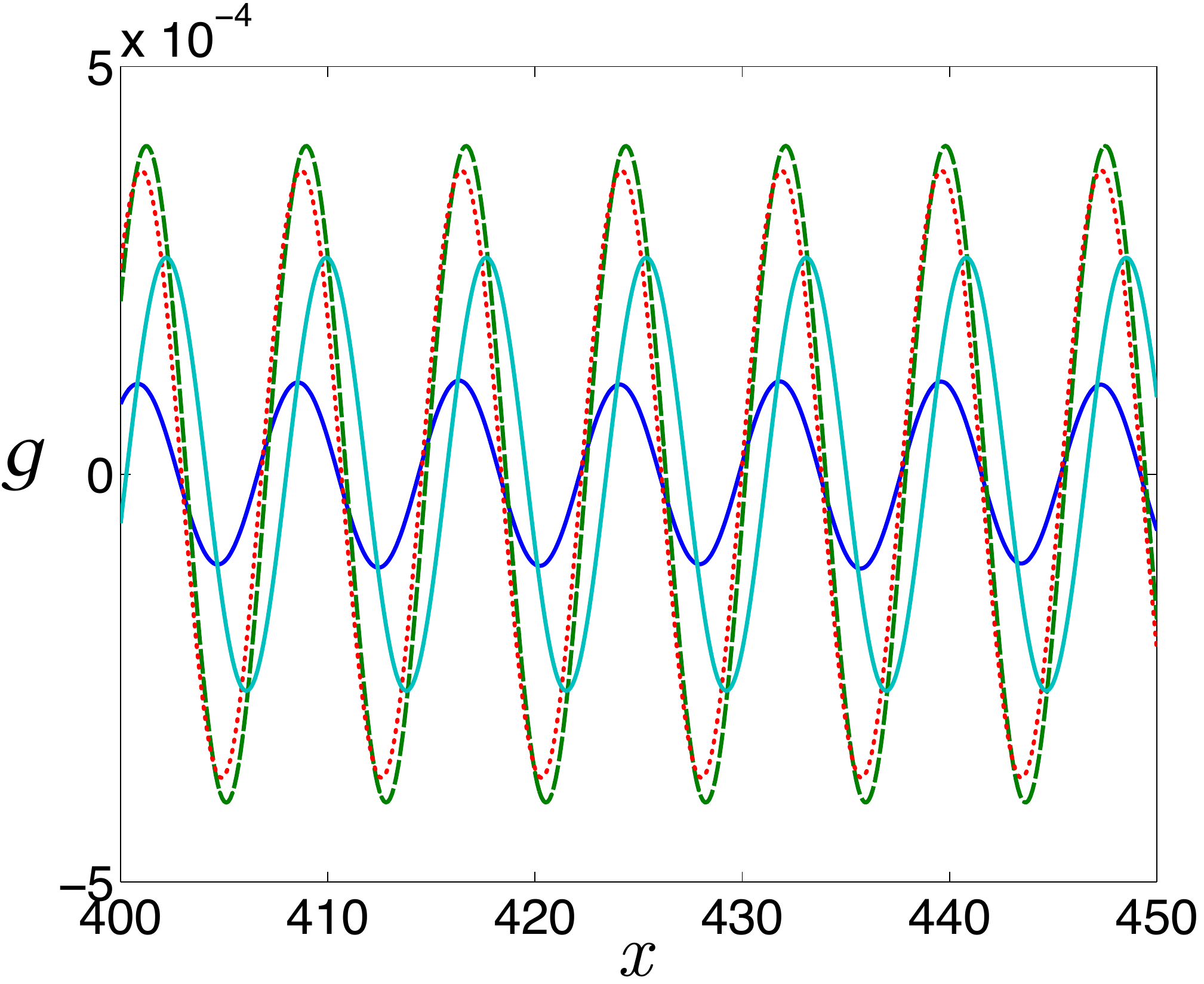}  \\
 \mbox{(c) \ \footnotesize \bf $g_{{\rm num}}$({\color{blue} \bf ---}) \  $g_{{\rm theory}}$({\color{KMgreen} \bf - -}) \ $g^1_{{\rm pert}}$$({\color{red} \bf \cdots})$ \  $g^2_{{\rm pert}}$({\color{Turquoise} \bf ---}) }   
 &  \mbox{(d) \ \footnotesize \bf $g_{{\rm num}}$({\color{blue} \bf ---}) \  $g_{{\rm theory}}$({\color{KMgreen} \bf - -}) \ $g^1_{{\rm pert}}$$({\color{red} \bf \cdots})$ \  $g^2_{{\rm pert}}$({\color{Turquoise} \bf ---}) }  
\end{array}$
\end{center}
\caption{\small Numerical solution and both theoretical solutions  at  $t=600$. Parameter values: $\varepsilon=0.35$, $k = 0.85$, $\mu=0.005$ and $A=7$, $B=C=0$, which implies $m = 0.8147$, $M = 0.8868$, $\omega=0.9576$, $c = 1.017$, $\alpha=\beta=1$, $\gamma=\delta = 0.297$, $v = 1.077$, $v_0 = 1.068$. Numerical parameters: $\Delta t = 0.01$, $L=2000$, $N=4\times10^5$.}
\label{figure:RSW_perturb_2nd}
\end{figure}
\begin{figure}[htbp]
\begin{center}$
\begin{array}{ccc}
\includegraphics[width=2.4in]{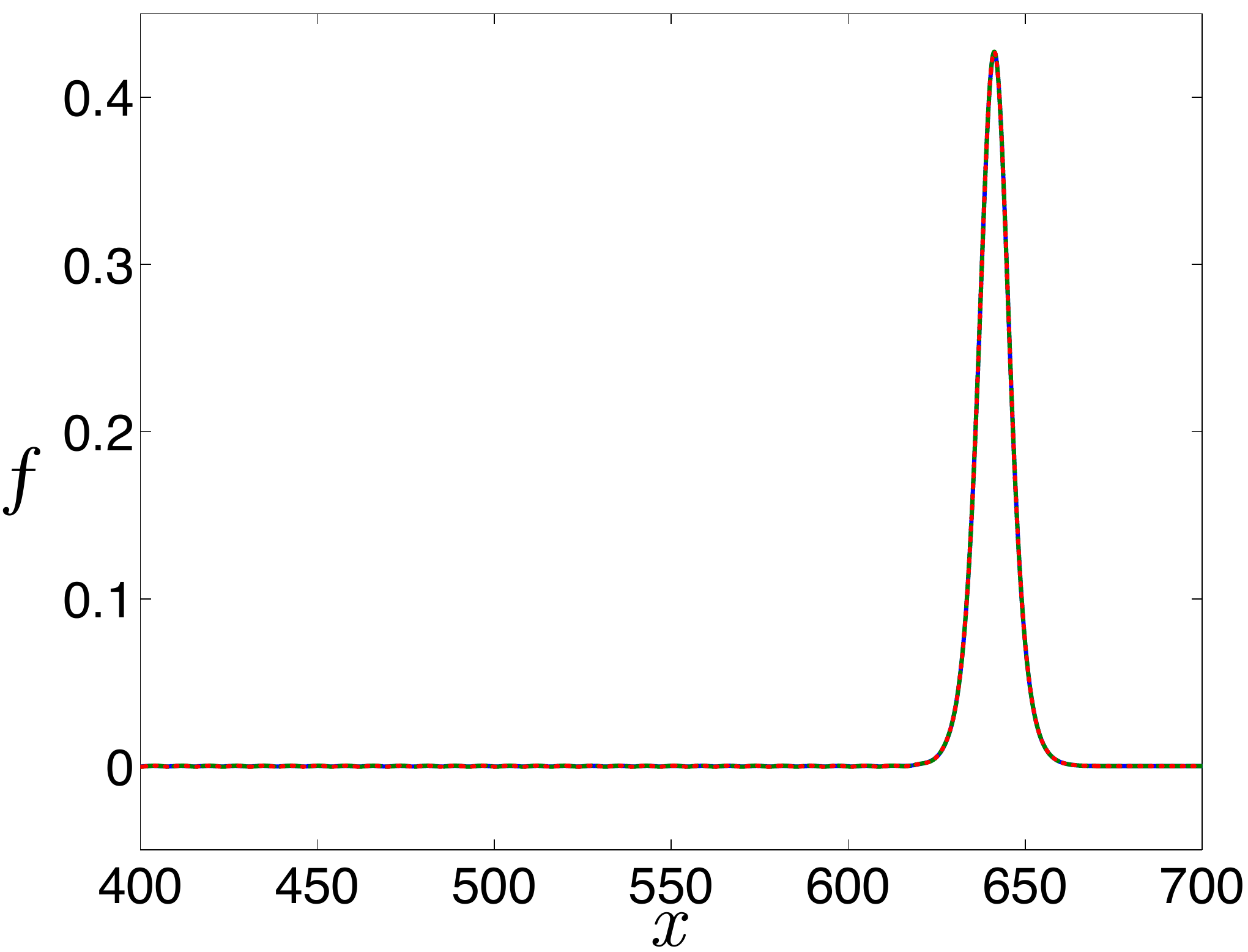}  &
\includegraphics[width=2.35in]{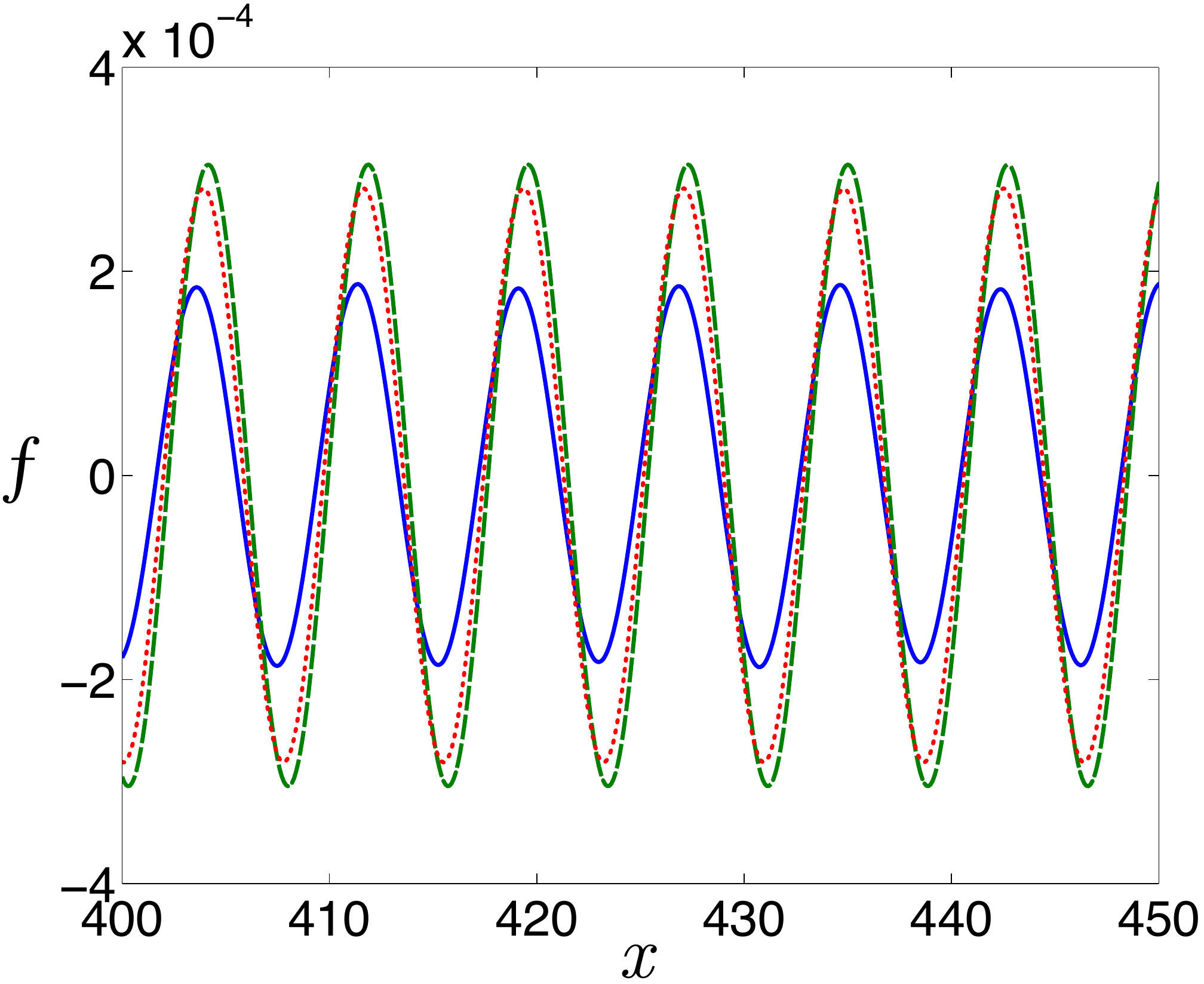}  \\
 \mbox{(a) \ \footnotesize \bf $f_{{\rm num}}$({\color{blue} \bf ---}) \  $f_{{\rm theory}}$({\color{KMgreen} \bf - -}) \ $f^2_{{\rm pert}}$$({\color{red} \bf \cdots})$ }   
 &  \mbox{(b) \ \footnotesize \bf $f_{{\rm num}}$({\color{blue} \bf ---}) \  $f_{{\rm theory}}$({\color{KMgreen} \bf - -}) \ $f^2_{{\rm pert}}$$({\color{red} \bf \cdots})$  }  
\\
\includegraphics[width=2.4in]{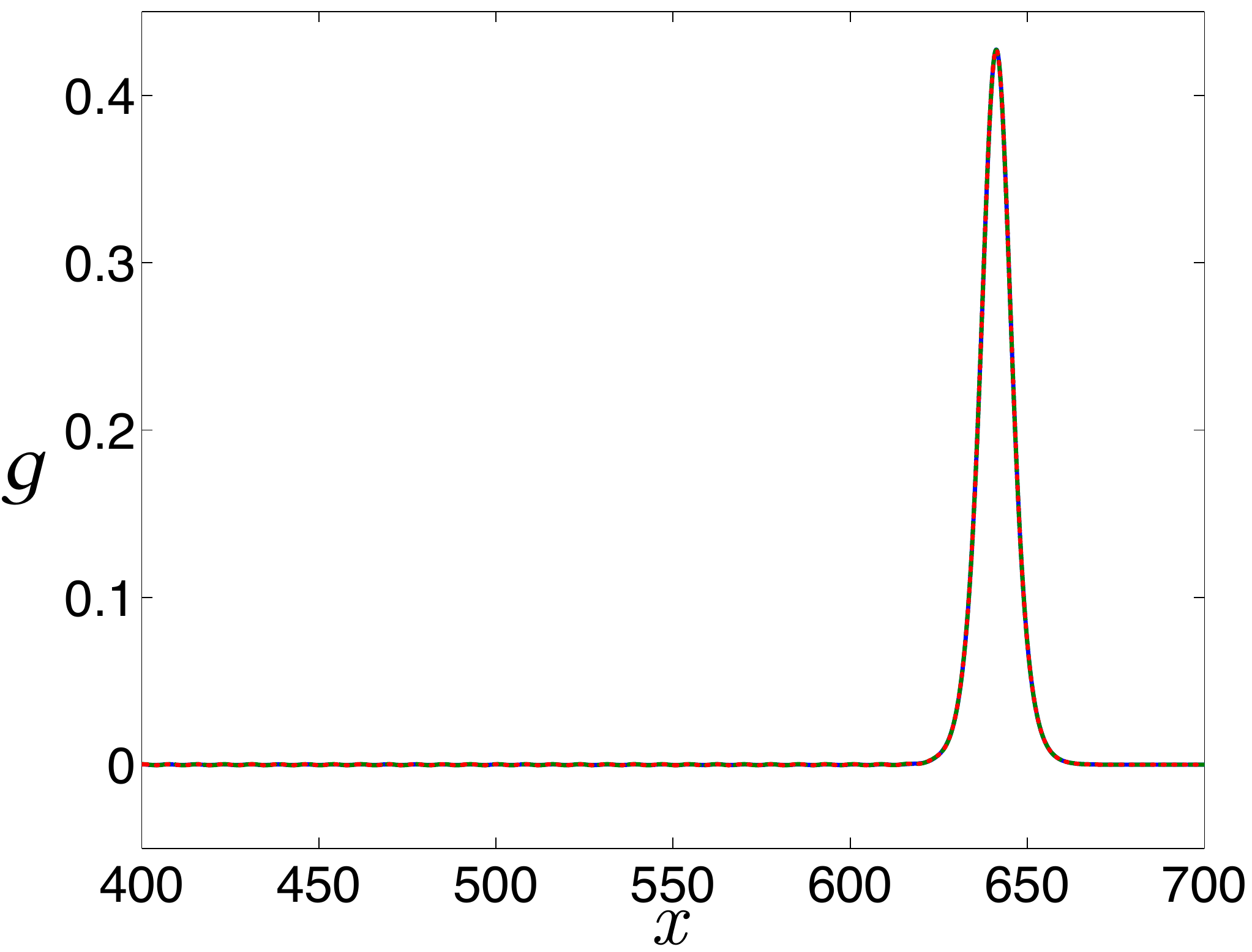}  &
\includegraphics[width=2.35in]{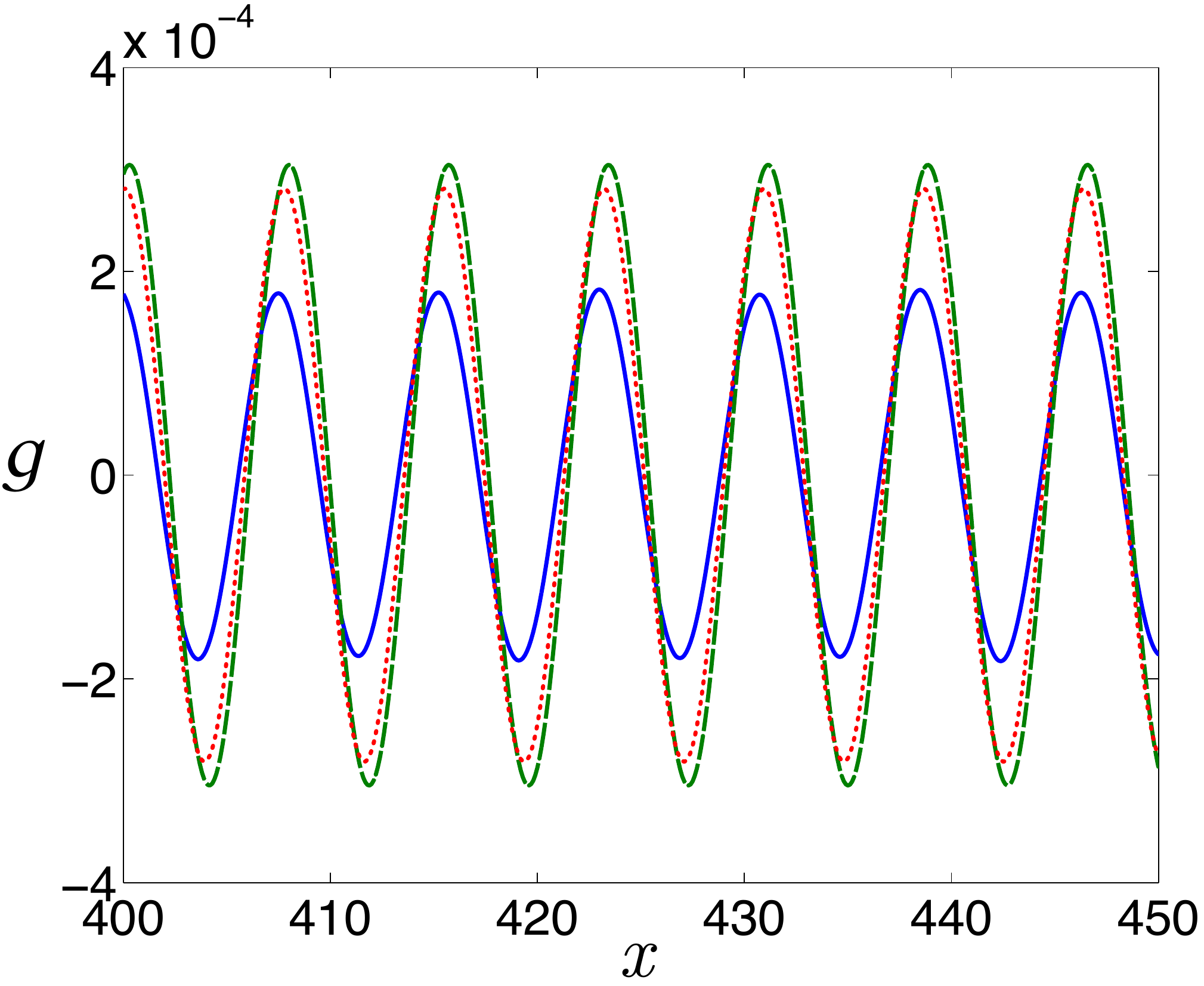}  \\
 \mbox{(c) \ \footnotesize \bf $g_{{\rm num}}$({\color{blue} \bf ---}) \  $g_{{\rm theory}}$({\color{KMgreen} \bf - -}) \ $g^2_{{\rm pert}}$$({\color{red} \bf \cdots})$ }   
 &  \mbox{(d) \ \footnotesize \bf $g_{{\rm num}}$({\color{blue} \bf ---}) \  $g_{{\rm theory}}$({\color{KMgreen} \bf - -}) \ $g^2_{{\rm pert}}$$({\color{red} \bf \cdots})$ }  
\end{array}$
\end{center}
\caption{\small Numerical solution and both theoretical solutions  at  $t=600$. Parameter values: $\varepsilon=0.35$, $k = 0.85$, $\mu=0.005$ and $A=B=0$, $C=7$, which implies $m = 0.8147$, $M = 0.8868$, $\omega=0.9576$, $c = \alpha=1$, $\beta=1.035$, $\gamma=\delta = 0.297$, $v = 1.069$, $v_0 = 1.068$. Numerical parameters: $\Delta t = 0.01$, $L=2000$, $N=4\times10^5$.}
\label{figure:RSW_perturb_beta_2nd}
\end{figure}

 \newpage
\section{Conclusion}

In this paper we have presented  two methods  to describe radiating solitary wave solutions inherent in  systems of coupled regularised Boussinesq (cRB) equations such as (\ref{fg}). Here we exploited the fact that when the 
system is symmetric ($c=\alpha =\beta = 1$ in (\ref{fg})) there is an exact solitary wave solution, 
and then the parameters in the system are slightly perturbed from the symmetric case, represented
by a parameter $\mu \ll 1$. 
This leads to a set of linearized equations (\ref{eqn:perturb_bous_sys_Omu}) which can be solved 
asymptotically using the method of variation of parameters. 
The solution represents a radiating solitary wave consisting of a 
localized central core and one-sided radiation. Our main interest is in the latter,  and to find explicit expressions for
the amplitude of the oscillations, we employed two different  approximation methods, both based on the assumption that
the amplitude of the leading order solitary wave is small, represented by $\varepsilon \ll 1$, 
see (\ref{eqn:soliton_later}). In one, described in section 2.2.1, we approximated the solutions of the homogeneous linearized  equation used in the method of variation of parameters by their far-field expressions, based on the fact that
a variable-coefficient term was $O(\varepsilon^2 )$.  
In the other, described in section 2.2.2, we used  the Lindstedt--Poincar\'e method in a 
systematic expansion in powers of $\varepsilon $.  In both methods the amplitude of the 
oscillations was exponentially small with respect to $\varepsilon $, and agreed in the limit
$\varepsilon \to 0$, although small but significant differences emerged for small but finite $\varepsilon$. 
We note that, although the radiation amplitudes are exponentially small, the present perturbation approach
avoids the necessity for the more complicated techniques of exponential asymptotics. However, as pointed out by 
\citet{akyyang} for instance, it is our  ordering that first $\mu \ll 1$ and then $\epsilon \to 0$ which avoids 
using exponential asymptotics, and a better procedure would be a simultaneous expansion in 
$\mu, \epsilon $. Nevertheless, our approach does yield the correct  wavenumber for the radiation, and a valid
estimate of the amplitude of the radiation which was confirmed by direct numerical simulations, and this is our main purpose here. 
We also note since the expressions we obtain for the radiation amplitudes depend linearly on three independent 
parameters $A, B, C$, see (\ref{eqn:E2_2nd_attempt}), (\ref{coeff-final}) for each method respectively, 
  there is a two-parameter family when $\varepsilon \to 0$ for which the amplitude of the radiating tail is zero. This
implies that then there is an embedded solitary wave inside the linear spectrum, see 
\citet{Ch, Yang} for instance.

In section 3 we compared  these theoretical results with some typical numerical simulations 
of the full system (\ref{fg}). The wavenumber 
of the oscillations was 
found to be accurately predicted by the theory, but there were discrepancies in the amplitude which was
larger in the theory than in the numerical simulations for the cases shown. 
Importantly we found that the Lindstedt--Poincar\'e method of section 2.2.2 gave better 
agreement than the {\it ad hoc} method of section 2.2.1. In both cases the discrepancies were
due to the finite value of $\varepsilon $ and not due to the truncation at $O(\mu)$. 
Also the { second} method has the advantage that
it can in principle be carried through to higher order in $\varepsilon $.

Recently, the scattering of a long longitudinal radiating bulk strain solitary wave in a delaminated area of a bi-layer with a soft adhesive bonding was modelled numerically by \cite{KT}. The problem was studied within the scope of the system (\ref{fg}), and the results indicated that radiating solitary waves could help us to control the integrity of layered structures with adhesive bondings. The generation of a radiating bulk strain solitary wave in a layered bar with a soft adhesive bonding was experimentally observed by \cite{DSSK}. The subsequent disappearance of the ``tail" in the delaminated area of the bi-layer was observed by \cite{Dreiden12}.  The estimates for the wavenumber and the amplitude of the ``tail" obtained in our paper will help to guide the subsequent numerical and laboratory experiments with adhesively bonded layered structures.


\bibliographystyle{apalike}

\end{document}